%% file: mdp.tex
\documentclass[12pt]{article}
\usepackage{amsmath}
\usepackage{texdraw}

\newcommand{\msbar}{$\overline{\mathrm MS}$}
\begin{document}
\title{\bf{A Lattice Study of Spectator Effects in Inclusive Decays of
$\mathbf{B}$-Mesons}}  
\author{\bf{UKQCD Collaboration}:\\
\bf{M. Di Pierro and C.T. Sachrajda}\\
\mbox{}\\  
Dept. of Physics and Astronomy,\\ 
University of Southampton,\\ 
Southampton SO17 1BJ, UK.}  
\date{} 
\maketitle

\vspace{-3.8in}
\begin{flushright}
SHEP 98-07\\ 
hep-lat/9805028
\end{flushright}
\vspace{2.9in}

\begin{abstract}
We compute the matrix elements of the operators which contribute to
spectator effects in inclusive decays of $B$-mesons. The results agree
well with estimates based on the vacuum saturation (factorization)
hypothesis. For the ratio of lifetimes of charged and neutral mesons
we find $\tau(B^-)/\tau(B_d)=1.03\pm 0.02\pm 0.03$, where the first
error represents the uncertainty in our evaluation of the matrix
elements, and the second is an estimate of the uncertainty due to the
fact that the Wilson coefficient functions have only been evaluated at
tree-level in perturbation theory. This result is in agreement with the
experimental measurement. We also discuss the implications of our
results for the semileptonic branching ratio and the charm yield.
\end{abstract}

\section{Introduction}
\label{sec:intro}

Inclusive decays of heavy hadrons can be studied in the framework of
the heavy quark expansion, in which widths and lifetimes are computed
as series in inverse powers of the mass of the
$b$-quark~\cite{chay,bsuv1,manohar} (for recent reviews and
additional references see refs.~\cite{reviews1,reviews2}). The leading
term of this expansion corresponds to the decay of a free-quark and is
universal, contributing equally to the lifetimes of all beauty
hadrons. Remarkably there are no corrections of $O(1/m_b)$, and the
first corrections are of $O(1/m_b^2)$~\cite{bsuv,manohar}. ``Spectator
effects", i.e. contributions from decays in which a light constituent
quark also participates in the weak process, enter at third order in
the heavy-quark expansion, i.e. at $O(1/m_b^3)$.  However, as a result
of the enhancement of the phase space for 2$\to$2 body reactions,
relative to 1$\to$3 body decays, the spectator effects are likely to
be larger than estimates based purely on power counting, and may well
be significant. The need to evaluate the spectator effects is
reinforced by the striking discrepancy between the experimental result
for the ratio of lifetimes~\cite{junk}:
\begin{equation}
\frac{\tau(\Lambda_b)}{\tau(B_d)}=0.78 \pm 0.04 \ ,
\label{eq:lbexp}\end{equation}
and the theoretical prediction
\begin{equation}
\frac{\tau(\Lambda_b)}{\tau(B_d)}=0.98+O(1/m_b^3)\ .
\label{eq:lbth}\end{equation}
In order to explain this discrepancy in the conventional approach, the
higher order terms in the heavy quark expansion, and the spectator
effects in particular, would have to be surprisingly large. In this
paper we compute the matrix elements of the four-quark operators
needed to evaluate these spectator effects for mesons; a computation
of the effects for the $\Lambda_b$ is in progress and the results will
be reported in a future publication.

The experimental value of the ratio of lifetimes of the charged and
neutral $B$-mesons is~\cite{junk}
\begin{equation}
\frac{\tau(B^-)}{\tau(B^0_d)}=1.06\pm 0.04\ ,
\label{eq:rtaubexp}\end{equation}
to be compared to the theoretical prediction
\begin{equation}
\frac{\tau(B^-)}{\tau(B^0_d)}=1 + O(1/m_b^3)\ .
\label{eq:rtaubth}\end{equation}
Below we determine the contribution to the $O(1/m_b^3)$ term on the
right-hand side of eq.(\ref{eq:rtaubth}) coming from spectator
effects, which we believe to be the largest component. The same
spectator effects also contribute to the semileptonic branching ratio
of the $B$-mesons, and to the charm-yield ($n_c$, the average number
of charmed particles in decays of $B$-mesons). This will be briefly
discussed in section~\ref{sec:concs}.

At $O(1/m_b^3)$, the non-perturbative contributions to the spectator
effects are contained in the matrix elements of the four-quark
operators:
\begin{eqnarray}
O_{V{-}A}^q & = & \bar b_L\gamma_\mu q_L\,\bar q_L\gamma^\mu b_L
\label{eq:ovadef}\\ 
O_{S{-}P}^q & = & \bar b_R q_L\,\bar q_L b_R
\label{eq:ospdef}\\ 
T_{V{-}A}^q & = & \bar b_L\gamma_\mu T^a q_L\,\bar q_L\gamma^\mu T^a b_L
\label{eq:tvadef}\\ 
T_{S{-}P}^q & = & \bar b_R T^a q_L\,\bar q_L T^a b_R\ ,
\label{eq:tspdef}
\end{eqnarray}
where $q$ represents the field of the light quark, and the $T^a$'s are
the generators of the colour group. Throughout this paper we take
these operators to be defined at a renormalization scale $m_b$. The
subscripts $L$ and $R$ represent ``left'' and ``right'' respectively.
For mesons, following ref.~\cite{ns}, we introduce the parameters
$B_{1,2}$ and $\varepsilon_{1,2}$:
\begin{eqnarray}
\frac{1}{2m_{B_q}}\,\langle\,B_q\,|\,O_{V{-}A}^q\,|\,B_q\rangle &
\equiv & \frac{f^2_{B_q}m_{B_q}}{8}\,B_1\ ,\label{eq:b1def}\\ 
\frac{1}{2m_{B_q}}\,\langle\,B_q\,|\,O_{S{-}P}^q\,|\,B_q\rangle &
\equiv & \frac{f^2_{B_q}m_{B_q}}{8}\,B_2\ ,\label{eq:b2def}\\ 
\frac{1}{2m_{B_q}}\,\langle\,B_q\,|\,T_{V{-}A}^q\,|\,B_q\rangle &
\equiv & \frac{f^2_{B_q}m_{B_q}}{8}\,\varepsilon_1\ ,\label{eq:eps1def}\\ 
\frac{1}{2m_{B_q}}\,\langle\,B_q\,|\,T_{S{-}P}^q\,|\,B_q\rangle &
\equiv & \frac{f^2_{B_q}m_{B_q}}{8}\,\varepsilon_2\ ,\label{eq:eps2def}
\end{eqnarray}
where $f_{B_q}$ is the leptonic decay constant of the meson $B_q$ (in
a normalization in which $f_\pi\simeq$~131~MeV). The introduction of
the parameters $B_{1,2}$ and $\varepsilon_{1,2}$ is motivated by the
vacuum saturation (or factorization) approximation~\cite{fact} in which
the matrix elements of four-quark operators are evaluated by inserting
the vacuum inside the current products. This leads to $B_i=1$ and
$\epsilon_i=0$ at some renormalization scale, $\mu$, for the
operators. It may be argued that the scale at which the factorization
approximation holds is different from our chosen renormalization scale
$m_b$, specifically that if the approximation is valid at all then it
should hold at a typical hadronic scale of about 1~GeV~\cite{blok}. In
QCD, in the limit of a large number of colours $N_c$,
\begin{equation}
\begin{array}{ccc}
B_i=O(1) & \textrm{and} & \varepsilon_i=O(1/N_c)\ .\\ 
\end{array}
\end{equation}
In this paper we evaluate the parameters $B_i$ and $\varepsilon_i$ in
lattice simulations, thus testing the validity of the factorization
hypothesis.

The results of our calculations indicate that the  vacuum saturation
hypothesis is (surprisingly?) well satisfied. In the \msbar\
renormalization scheme we find:
\begin{alignat}{2}
B_1(m_b) &=1.06(8) & \qquad
B_2(m_b) &=1.01(6) \label{eq:bmsbar0}\\
\varepsilon_1(m_b) &=-0.01(3) & \qquad
\varepsilon_2(m_b) &=-0.01(2) \label{eq:epsmsbar0}\
\end{alignat}
and
\begin{equation}
\frac{\tau(B^-)}{\tau(B_d)} = 1.03\pm 0.02\pm 0.03\ ,
\label{eq:result}\end{equation}
in good agreement with the experimental value in
eq.~(\ref{eq:rtaubexp}).

The present calculation is very similar to that of the $B_B$-parameter
of $B$-$\bar B$ mixing, for which several recent simulations have been
performed~\cite{ewing,gm,draper}, including one using the same
configurations used in this study~\cite{ewing}. We use the
calculations of the $B_B$ parameter as a comparison and check on our
results, both for the perturbative matching coefficients and for the
evaluation of the matrix elements. In performing this comparison we
believe that we have found an error in the perturbative
calculation. We also stress that the feature that the values of the
matrix elements of the operators (\ref{eq:ovadef})-(\ref{eq:tspdef})
are close to those expected from the vacuum saturation hypothesis is
also present in the evaluation of the $B_B$-parameter.

The plan of the remainder of this paper is as follows. In the next
section we derive the relation, at one-loop order of perturbation
theory, between the operators defined in
eqs.~(\ref{eq:ovadef})-(\ref{eq:tspdef}) and the bare lattice
operators whose matrix elements are computed directly in lattice
simulations.  The technical details of the calculation are relegated
to appendix A. In section~\ref{sec:computation} we present a
description of the lattice computation, the results of this
computation and a discussion of the implications. The comparison of
our study to that of $B^0$-$\bar B^0$ mixing is presented in
section~\ref{sec:mixing} (the discussion of the evaluation of the
matching coefficients for $B$-$\bar B$ mixing is described in detail
in appendix B). Finally, in section~\ref{sec:concs} we present our
conclusions.

\section{Perturbative Matching}
\label{sec:matching}

The Wilson coefficient functions of the operators
(\ref{eq:ovadef})--(\ref{eq:tspdef}) in the OPE for inclusive decay
rates have been evaluated only at tree-level~\cite{ns}. At this level
of precision it is sufficient to compute the matrix elements in any
``reasonable'' renormalization scheme. In this paper we present the
results for the matrix elements of the HQET operators in the
$\overline{\mathrm MS}$ scheme at a renormalization scale
$m_b$~\footnote{From these it is possible to obtain the matrix
elements in any other renormalization scheme using perturbation
theory.}. In order to obtain these from the matrix elements of bare
operators in the lattice theory, with cut-off $a^{-1}$ (where $a$ is
the lattice spacing), which we compute directly in our simulations, we
require the corresponding matching coefficients. In this section we
present these matching coefficients (at one-loop order in perturbation
theory); we postpone a detailed description of their evaluation to
Appendix A.  In the following section we will use these coefficients
and the computed values of the matrix elements of the bare operators
to determine the coefficients $B_i$ and $\varepsilon_i$ in the
$\overline{\mathrm MS}$ scheme.

The coefficients presented in the section were all obtained using 
local lattice operators defined in the tree-level improved 
SW-action~\cite{sw} defined in eq.~(\ref{eq:sswdef}) below.

It is convenient to perform the matching in two steps:
\begin{enumerate}
\item[i)] The first step is the evaluation of the 
coefficients which relate the
HQET operators in the continuum ($\overline{\mathrm MS}$) and lattice
schemes, both defined at the scale $a^{-1}$,
\begin{equation}
O_i^{C}(a^{-1}) = O_i^{L}(a^{-1})  +
\frac{\alpha_s}{4\pi} D_{ij} O_j^{L}(a^{-1})\ ,
\label{eq:matching2}\end{equation}
where the lattice (continuum) operators are labelled by the superfix
$L$ ($C$). The mixing of operators is such that it is necessary to
compute the matrix elements of eight lattice operators
$O_j^{L}(a^{-1})$ (see table~\ref{tab:dij4}, where the coefficients
$D_{ij}$ are presented).
\item[ii)] We then evolve the HQET operators in the continuum scheme
from renormalization scale $\mu=a^{-1}$ to scale $\mu=m_b$, 
\begin{equation}
O_i^{C}(m_b) = M_{ij}(a^{-1},m_b)O_j^{C}(a^{-1})\ .
\label{eq:hybrid}\end{equation}
This evolution, sometimes known as ``hybrid renormalization", requires
knowledge of the anomalous dimension matrix~\cite{shif1,shif2,politzer}.
\end{enumerate}
\par The matching procedure involves short-distance physics only, and so
can be carried out in perturbation theory. Lattice perturbation theory
generally converges very slowly, and therefore, where possible, the
matching should be performed non-perturbatively so as to minimise these
systematic errors. In this letter, however, we do not use a
non-perturbative renormalization, but, when evaluating the matching
coefficients, we do use a ``boosted'' lattice coupling constant in order
to partially resum the large higher order contributions (e.g. those
coming from tadpole graphs).
\begin{table}
\begin{center}
\begin{tabular}{|c|c|c|c|c|c|}\hline
$D_{ij}$ & $i=1$ & $i=2$ & $i=3$ & $i=4$ & $O_j^{L}$
\rule[-8pt]{0pt}{22pt}\\ \hline
$j=1$ & {-}21.64 & - & 2.06 & 0.54 & 
$\bar b_L \gamma^\mu q_L\,\bar q_L \gamma^\mu b_L$
\rule{0pt}{14pt}\\ 
$j=2$ & - & {-}21.64 & 2.16 & 2.40 &  
$\bar b_R q_L\,\bar q_L b_R$ \rule{0pt}{14pt}\\ 
$j=3$ & 9.29 & 2.43 & {-}10.80 & 2.83 & 
$\bar b_L \gamma^\mu T^a q_L\,\bar q_L \gamma^\mu T^a b_L$
\rule{0pt}{14pt}\\ 
$j=4$ & 9.72 & 10.79 & 11.34 & {-}9.05 & 
$\bar b_R T^a q_L\,\bar q_L T^a b_R$
\rule{0pt}{14pt}\\  
$j=5$ & {-}18.37 & - & {-}3.06 & - & 
$\bar b_R \gamma^\mu q_R\,\bar q_L \gamma^\mu b_L$
\rule{0pt}{14pt}\\ 
$j=6$ & 36.75 & 18.37 & 6.12 & 3.06 &  
$\bar b_R q_L\,\bar q_R b_L$ \rule{0pt}{14pt}\\ 
$j=7$ & -13.78 & - & 6.89 & - & 
$\bar b_R \gamma^\mu T^a q_R\,\bar q_L \gamma^\mu T^a b_L$
\rule{0pt}{14pt}\\ 
$j=8$ & 27.56 & 13.78 & {-}13.78 & {-}6.89 & 
$\bar b_R T^a q_L\,\bar q_R T^a b_L$
\rule{0pt}{14pt}\\  \hline
\end{tabular}
\caption{Coefficients $D_{ij}$ for the four operators of
eqs.(\ref{eq:ovadef})-(\ref{eq:tspdef}).\label{tab:dij4}} \end{center}
\end{table}

In order to obtain the parameters $B_i$ and $\varepsilon_i$ it is also
necessary to determine the normalization of the axial current~\cite{bp}.
In this case there is a single coefficient $Z_A$ defined by 
\begin{equation}
\left\langle 0\right| A_0^C(a^{-1})\left| B\right\rangle=
Z_A\left\langle 0\right| A_0^L(a^{-1})\left| B\right\rangle\ ,
\label{eq:zadef}\end{equation}
which at one loop order in perturbation theory is given by:
\begin{equation}
Z_A=1+\frac{\alpha _s(a^{-1})}{4\pi }\frac 43\left( \frac
54-\frac12 x_1-x_2+x_3\right) \simeq 1 - 
20.0\,\frac{\alpha _s(a^{-1})}{4\pi }=0.79  \,
\label{eq:zanum}\end{equation}
where the coefficients $x_i$ are defined in eq.~(\ref{eq:dici}) and
tabulated in table~\ref{tab:cis} in Appendix A. $A_0^C$ and $A_0^L$
are the time components on the axial current in the HQET in the
$\overline{\mathrm MS}$ and lattice schemes respectively, both defined
at the scale $a^{-1}$. Using hybrid renormalization one can obtain the
axial current in the $\overline{\mathrm MS}$ scheme at scale $m_b$.
We stress again that the results presented for the $B_i$'s and
$\varepsilon_i$'s below were obtained with both the four quark
operators and the axial current defined in the HQET in the 
$\overline{\mathrm MS}$ scheme at the scale $m_b$~\footnote
{These differ at one-loop level from the corresponding operators
in QCD.}.

In the following section we combine the results of the matrix elements
computed on the lattice with the perturbative coefficients presented
in this section to obtain the $B_i$'s and $\varepsilon_i$'s.

\section{Lattice Computation and Results}
\label{sec:computation}

The non-perturbative strong interaction effects in spectator
contributions to inclusive decays are contained in the matrix elements
of the eight four-quark operators, $O_j$, given in table~\ref{tab:dij4}.
We evaluate these matrix elements in a quenched simulation on a 
$24^3\times 48$ lattice at
$\beta=6.2$ using the $SW$ tree-level improved action~\cite{sw}:
\begin{equation}
S^{SW}=S^{\rm{gauge}} + S^{\rm{Wilson}} - i\frac{\kappa}{2}
\sum_{x,\mu,\nu}\bar q(x)F_{\mu\nu}(x)\sigma_{\mu\nu}\,q(x)\ ,
\label{eq:sswdef}\end{equation}
where $S^{\rm{gauge}}$ and $S^{\rm{Wilson}}$ are the Wilson gauge and
quark actions respectively. The use of this action reduces the errors
due to the granularity of the lattice to ones of $O(\alpha_s a)$, where
$a$ is the lattice spacing. We use the 60 SU(3) gauge-field
configurations, and the light quark propagators corresponding to
hopping parameters $\kappa =\,$0.14144, 0.14226 and 0.14262 which have
been used previously to obtain the $B$-parameter of $B^0$-$\bar B^0$
mixing~\cite{ewing} and other quantities required for studies of
$B$-physics.  The value of the hopping parameter in the chiral limit
is given by $\kappa_c=0.14315$. The calculation of the matrix elements
of the operators in table~\ref{tab:dij4} is very similar to that of
the $\Delta B=2$ operators from which the $B$-parameter is extracted,
and we exploit the similarities to perform checks on our calculations
(see section~\ref{sec:mixing} and Appendix B).

The evaluation of the matrix elements requires the computation of
two- and three-point correlation functions:
\begin{equation}
C(t_x)\equiv\sum_{\vec x}\langle\,0\,|
\,J(x)J^\dagger(0)\,|\,0\,\rangle\ ,
\label{eq:cdef}\end{equation}
where we have assumed that $t_x>0$, and 
\begin{equation}
K(t_x,t_y)\equiv\sum_{\vec x,\vec y}\langle\,0\,|
\,J(y)O_j^{L}(0)J^\dagger(x)\,|\,0\,\rangle\ ,
\label{eq:kdef}\end{equation}
where $t_y>0>t_x$. In eqs.~(\ref{eq:cdef}) and (\ref{eq:kdef})
$J^\dagger$ and $J$ are interpolating operators which can create or
destroy a heavy pseudoscalar meson (containing a static heavy quark)
and $O_j$ is one of the operators whose matrix element we wish to
evaluate. In practice we choose J to be the fourth component of an
axial current. It is generally advantageous to ``smear'' the
interpolating operators $J$ and $J^\dagger$ over the spatial
coordinates, in order to enhance the overlap with the ground
state. Following ref.~\cite{ewing} we have used several methods of
smearing and checked that the results for the matrix elements of $O_j$
are independent of the choice of smearing. In this paper we present as
our ``best'' results those obtained using a gauge-invariant smearing
based on the Jacobi algorithm described in
ref.~\cite{jacobi}. The smeared heavy-quark field at time $t$,
$b^S(\vec x,t)$, is defined by
\begin{equation}
b^S(x) = \sum_{\vec{x^\prime}} M(\vec x,\vec x^\prime)b(\vec
x^\prime,t)\ ,
\label{eq:jdef}\end{equation}
where 
\begin{equation}
M(\vec x,\vec x^\prime) = \sum_{n=0}^{N}\kappa_S\Delta^n(\vec x,\vec
x^\prime)
\label{eq:mdef}\end{equation}
and $\Delta$ is the three-dimensional version of the Laplace operator.
The parameter $\kappa_S$ and the number of iterations can be used to
control the smearing radius, and here we present our results for
$\kappa_s=0.25$ and $N=140$ which corresponds to an rms smearing
radius $r_0=6.4a$~\cite{smearing}. The smeared interpolating operator is then
chosen to be $J^S(x)=\bar b^S(x)\gamma_5q(x)$, where $q$ is the field of
the light quark. The local interpolating operator is simply defined to be
$J^L(x)=\bar b(x)\gamma_5q(x)$.

The eight matrix elements $\langle B|\,O_j^{L}(0)\,|B\rangle$ are
determined by computing the ratios:
\begin{equation}
R_j(t_1,t_2)\equiv\frac{4K_j^{SS}(-t_1,t_2)}{C^{LS}(-t_1)C^{LS}(t_2)}\ ,
\label{eq:rjdef}\end{equation}
where $t_1$ and $t_2$ are positive and sufficiently large to ensure that
only the ground state contributes to the correlation functions. The
indices $S$ and $L$ denote whether the interpolating operators are
smeared or local. It is convenient to choose both $J$ and $J^\dagger$ to
be smeared in the three-point function $K_j$ and to evaluate the
two-point functions with a local operator at the source and a smeared
one at the sink. At large time separations $t_1$ and $t_2$,
\begin{equation}
R_j(t_1,t_2)\to\frac{2}{m_BZ_L^2}
\langle B|\,O_j^{L}(0)\,|B\rangle\ ,
\label{eq:rjasymp}\end{equation}
where $Z_L$ is given by
\begin{equation}
\langle0|\,J^L(0)\,|B\rangle\equiv\sqrt{2m_B}Z_L\ .
\label{eq:zldef}\end{equation}

The $R_j$ defined in eq.(\ref{eq:rjdef}) contribute directly to the
$B_i$'s and the $\varepsilon_i$'s, apart from perturbative matching
factors. To see this, note that the leptonic decay constant of the
$B$-meson in the static limit (i.e. infinite mass limit for the
$b$-quark) is given by
\begin{equation}
f_B^2=\frac{2Z_L^2Z_A^2}{m_B}\ ,
\label{eq:fbzl}\end{equation}
so that
\begin{equation}
R_j(t_1,t_2)\to\frac{2}{m_BZ_L^2}\langle B|\,O_j^{L}(0)\,|B\rangle
=\frac{4Z_A^2}{f_B^2m_B^2}
\langle B|\,O_j^{L}(0)\,|B\rangle\ ,
\label{eq:rj2}\end{equation}
which corresponds precisely to the normalization of the $B_i$'s and
$\varepsilon_i$'s in eqs.(\ref{eq:b1def})--(\ref{eq:eps2def}) (apart
from the matching factors $Z_A^2$ and that of the
four-quark operator $O_j$ described in section~\ref{sec:matching}).

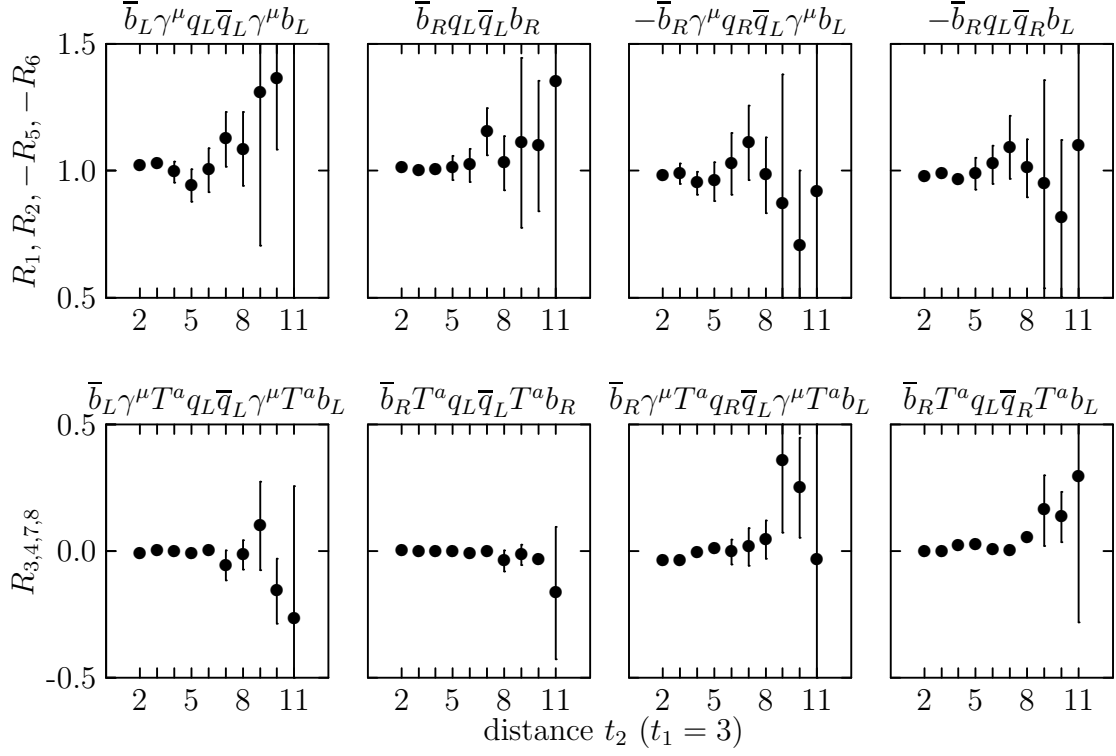
\begin{figure}[t]
\input{mydraw.tex}
\caption{Results for the $R_j,\ j=1\cdots 8$ as a function of $t_2$
for $t_1=3$. The value of the quark mass corresponds to $\kappa
=0.14226$.\label{fig:rjresults}}
\end{figure}

In these computations it is particularly important to establish that
the contribution from the ground-state has been isolated. The natural
way to do this is to look for plateaus, i.e. for regions in $t_1$ and
$t_2$ for which $R_j(t_1,t_2)$ is independent of $t_1$ and
$t_2$. Since the statistical errors grow fairly quickly with $t_{1,2}$
our ability to verify the existence of plateaus is limited. For
example in fig.~\ref{fig:rjresults} we present our results for
$R_j(t_1,t_2)$, obtained with the light quark mass corresponding to
$\kappa=0.14226$ as a function of $t_2$ for $t_1=3$.  The results are
consistent with being constant, but the errors are uncomfortably large
for $t_2>5$ or so. An alternative, and perhaps more convincing, way to
confirm that the contribution from the ground-state has been isolated
is to check that the values of the $R_j$ are independent of the method
of smearing used to define the smeared interpolating
operators. Following ref.~\cite{ewing}, in addition to the gauge
invariant prescription for smearing described above, we have defined
the smeared field in four other ways (having transformed the fields to
the Coulomb gauge). For these additional four cases, 
$M(\vec x,\vec x')$ of eq.~(\ref{eq:mdef}) is replaced by the following:
\begin{alignat}{2}
\textrm{Exponential}: & \qquad M(\vec x,\vec x')& \,=\, &\exp(-|\vec x-\vec x'|/r_0),
\label{eq:smexp}\\ 
\textrm{Gaussian}: & \qquad M(\vec x,\vec x')& \,=\, &\exp(-|\vec x-\vec x'|^2/r_0^2), 
\label{eq:smgauss}\\
\textrm{Cube}: & \qquad M(\vec x,\vec x')& \,=\, &\prod_{i=1}^3\Theta(r_0 -
|\vec x_i-x'_i|), 
\label{eq:smcube}\\
\textrm{Double Cube}: & \qquad M(\vec x,\vec x')& \,=\, &\prod_{i=1}^3
\left(1-\frac{|x_i-x'_i|}{2r_0}\right)
\Theta(2r_0 -|x_i- x'_i|), 
\label{eq:smdcube}
\end{alignat}
and we have chosen $r_0=5$. As an example, we present in
table~\ref{tab:rjsm} the results for the $R_j(t_1,t_2)$ for $t_1=3$
and $t_2=3$, again for the middle value of the three $\kappa$'s,
$\kappa=0.14226$.

\begin{table}
\begin{center}
\begin{tabular}{|c|c|c|c|c|c|}\hline
 & Gauge Inv. & Exponential & Gaussian & Cube & Double Cube \\ \hline 
$R_1(3,3)$ &  $1.02(3)$ &  $1.03(2)$ & $1.04(3)$  & $1.03(2)$  & $ 1.07(8)$ \\ 
$R_2(3,3)$ &  $1.00(2)$ & $1.01(2)$  & $1.01(3)$  & $1.01(2)$  & $1.00(5)$ \\ 
$R_3(3,3)$ &  $0.00(1)$ & $-0.00(1)$ & $-0.00(2)$ & $-0.00(1)$ & $-0.03(3)$ \\ 
$R_4(3,3)$ &  $0.00(1)$ &  $0.00(1)$ & $0.01(1)$  & $0.00(1)$  & $0.01(2)$ \\ 
$R_5(3,3)$ & $-0.99(4)$ & $-1.01(3)$ & $-1.01$(4) & $-1.00(3)$ & $-1.01(7)$ \\ 
$R_6(3,3)$ & $-0.99(3)$ & $-1.00(2)$ & $-1.01(3)$ & $-1.00(3)$ & $-1.01(6)$ \\ 
$R_7(3,3)$ & $-0.04(2)$ & $-0.03(2)$ & $-0.04(2)$ & $-0.03(2)$ & $-0.09(5)$ \\ 
$R_8(3,3)$ & $-0.00(1)$ & $-0.00(1)$ & $-0.01(1)$ & $-0.01(1)$ & $-0.02(2)$ \\ 
\hline 
\end{tabular}
\end{center}
\caption{Values of the $R_j(3,3)$ obtained with different smearing methods
for a light quark with $\kappa=0.14226$.\label{tab:rjsm}} 
\end{table}

We take as our best results those obtained with the gauge invariant
smearing, and with $t_1=t_2=3$. After extrapolating to the chiral
limit we find the following results for the $R_j$:
\begin{alignat}{2}
R_1(3,3)&= 1.04 \pm 0.04 & \qquad 
R_2(3,3)&= 1.00 \pm 0.03 \label{eq:r12res}\\  
R_3(3,3)&= -0.01 \pm 0.02 &\qquad 
R_4(3,3)&= -0.00 \pm 0.01 \label{eq:r34res}\\  
R_5(3,3)&= -0.97 \pm 0.06 &\qquad 
R_6(3,3)&= -0.98 \pm 0.05 \label{eq:r56res}\\  
R_7(3,3)&= -0.03 \pm 0.03 &\qquad 
R_8(3,3)&= -0.01 \pm 0.01 \label{eq:r78res}
\end{alignat}

We now combine these results with the matching coefficients presented in
section~\ref{sec:matching} to determine the $B_i$'s and the
$\varepsilon_i$'s. For the value of the lattice coupling constant we take
one of the  standard definitions of the boosted coupling:
\begin{equation}
\frac{\alpha_s(a^{-1})}{4 \pi}=\frac{6 (8 \kappa_{c})^4}{(4
\pi)^2 \beta} =0.0105\ .
\label{eq:boosted}\end{equation}
Using the coefficients from section~\ref{sec:matching} we then obtain
\begin{eqnarray*}
B_1(a^{-1}) &=&\left( Z_A\right) ^{-2}
\left( 1+\frac{\alpha _s(a^{-1})}{4\pi }D\right)_{1j}R_j 
= 0.98(8) \\
B_2(a^{-1}) &=&\left( Z_A\right) ^{-2}
\left( 1+\frac{\alpha _s(a^{-1})}{4\pi }D\right)_{2j}R_j 
= 0.93(5)\\
\varepsilon _1(a^{-1}) &=&\left( Z_A\right) ^{-2}
\left( 1+\frac{\alpha _s(a^{-1})}{4\pi }D\right)_{3j}R_j 
= 0.01(3)\\
\varepsilon _2(a^{-1}) &=&\left( Z_A\right) ^{-2}
\left( 1+\frac{\alpha _s(a^{-1})}{4\pi }D\right)_{4j}R_j 
= 0.00(2) \
\end{eqnarray*}

We now evolve those coefficients to the scale $m_b$, using the 
relations: 
\begin{eqnarray}
B(m_b) &=&\left[1+\frac{2C_F\delta}{N_c}\right]B(a^{-1})
	  -\frac{2\delta}{N_c}\varepsilon(a^{-1}) \\
\varepsilon(m_b) &=&\left[1+\frac{\delta}{N_c^2}\right]\varepsilon(a^{-1})
	  -\frac{C_F\delta}{N_c^2}B(a^{-1}) \ ,
\end{eqnarray}
where 
\begin{equation}
\delta\equiv\left(\frac{\alpha _s(a^{-1})}{\alpha _s(m_b)}\right) ^{9/2\beta
_0}-1=0.09(3)\ .
\end{equation}

In estimating $\delta$ and its uncertainty we have allowed for a
conservative variation of  the parameters around the ``central'' values
($\Lambda_{QCD}=250\,$MeV, $a^{-1}=2.9\,$GeV, $m_b=4.5\,$GeV and
$\beta_0 =9$). We finally obtain:
\begin{alignat}{2}
B_1(m_b) &=1.06(8) & \qquad
B_2(m_b) &=1.01(6) \label{eq:bmsbar}\\
\varepsilon_1(m_b) &={-}0.01(3) & \qquad
\varepsilon_2(m_b) &={-}0.01(2) \ .\label{eq:epsmsbar}\
\end{alignat}

These matrix elements have also been evaluated using QCD 
sum-rules~\cite{sumrules}. These authors find
$B_1(m_b)=0.96(4),\,B_2(m_b)=0.95(2),\,\varepsilon_1(m_b)=-0.14(1)$
and $\varepsilon_2(m_b)=-0.08(1)$, differing somewhat (particularly for the 
$\varepsilon_i$'s) from our values. 

Using the results in eqs.~(\ref{eq:bmsbar}) and (\ref{eq:epsmsbar}) 
we obtain the following value for the ratio of
lifetimes for the neutral and charged $B$-mesons: 
\begin{equation}
\frac{\tau (B^{-})}{\tau (B_d)}=1+k_1B_1+k_2B_2+k_3\varepsilon
_1+k_4\varepsilon _2= 1.03\pm 0.02\pm 0.03
\label{eq:ratiovalue}\end{equation}
where the coefficients $k_i$ are taken from ref.~\cite{ns}. The first
error in eq.~(\ref{eq:ratiovalue}) is from the uncertainty in the values
of the matrix elements in eqs.~(\ref{eq:bmsbar}) and
(\ref{eq:epsmsbar}), whereas the second is an estimate of the
uncertainty due to our ignorance of the one-loop contribution to the
Wilson coefficient function in the OPE (this was estimated by varying
the matching scale from $m_b/2$ to $2 m_b$ using the procedure described
in ref.~\cite{ns}). The value in eq.~(\ref{eq:ratiovalue}) is in good
agreement with the experimental results in eq.~(\ref{eq:rtaubexp}). 

There has been a considerable amount of discussion lately as to whether
the theoretical predictions for the semileptonic branching ratio of the
$B$-meson ($B_{SL}$) and the charm yield are consistent with
experimental measurements~\cite{ap,baffling,ns}. Here we limit our
discussion to a comment on the implications of our results on these two
physical quantities. The spectator contributions to $B_{SL}$ and $n_c$
take the form:
\begin{eqnarray}
\Delta B_{SL,\textrm{spec}} & = & b_1B_1+ b_2 B_2 + b_3 \varepsilon_1
+ b_4 \varepsilon_2\label{eq:deltaB}\\
\Delta n_{c,\textrm{spec}} & = & n_1B_1+ n_2 B_2 + n_3 \varepsilon_1
+ n_4 \varepsilon_2\label{eq:nc}\ ,
\end{eqnarray}
where the coefficients $b_i$ and $n_i$ at tree-level are presented
explicitly in ref.~\cite{ns} (see also
refs.~\cite{blok,guberina,shif1}).  The uncertainties in the
predictions for $B_{SL}$ and $n_c$ due to the errors in the matrix
elements in eqs.~(\ref{eq:bmsbar}) and (\ref{eq:epsmsbar}) are small
(about 0.1\% for both $B_{SL}$ and $n_c$). Larger uncertainties are
due to our ignorance of the one-loop contribution to the Wilson
coefficient function in the OPE. For example, following the procedure
described in ref.~\cite{ns} with a matching scale $\mu = m_b/2$ we
find $\Delta B_{SL,\textrm{spec}}=0.3(1)\%$ and $\Delta
n_{c,\textrm{spec}} =0.5(2)\%$, whereas for a scale $2m_B$ the
corresponding results are $\Delta B_{SL,\textrm{spec}}=-0.1(1)\%$ and
$\Delta n_{c,\textrm{spec}} =0.0(1)\%$. For the charm yield these
corrections are negligible as expected, and for the semileptonic
branching ratio they are small.  Nevertheless, it would be useful to
know the one-loop contribution to the coefficient functions in order
to eliminate the variation obtained by changing the matching scale.

\section{$B$-$\bar B$ Mixing}
\label{sec:mixing}

In this section we revisit the phenomenologically important process
of $B$-$\bar B$ mixing. The computation of the matrix elements of the 
relevant lattice operators on the same gauge configurations has already
been presented in ref.~\cite{ewing}, and we do not add to these lattice
computations. We do, however, wish to make two observations:
\begin{enumerate}
\item[i)] The matrix elements of the lattice $\Delta B=2$ operators
relevant for $B$-$\bar B$ mixing factorize with a similar precision to
that found for the four-quark operators considered in
section~\ref{sec:computation}. Because of the way in which lattice
computations are usually organised, this property is not as readily
manifest for the $\Delta B$=2 operators as it is for the spectator
effects.  However, by using colour and spin Fierz identities, we
demonstrate that the values of the matrix elements of the lattice
$\Delta B=2$ operators are very close to those expected using
factorization. Some of our observations have already been noted in
ref.~\cite{draper}, where the Wilson formulation for the light quarks
was used. We extend this investigation of factorization, and show that
each contribution to the matrix elements (i.e. each Wick contraction) is
close to the estimated value obtained using the factorization and
vacuum saturation hypothesis.
\item[ii)] We believe that there is an error in the published value
of the matching factors for the $\Delta B=2$ operators using the SW action
for the light quarks. We discuss this in some detail in Appendix B; in this
section we briefly comment on the consequences of this error. 
\end{enumerate}
We now consider these two observations in turn.

\subsection{Factorization}
\label{subsec:bbarfact}

In order to determine the $B$-parameter of $B$-$\bar B$ mixing we need to
evaluate the matrix elements of the three lattice operators $O_L,\,O_R$
and $O_N$ defined in eqs.(\ref{eq:oldef})--(\ref{eq:ondef}) of Appendix
B, as well as that of
\begin{equation}
O_S = \bar b q_L\,\bar b q_L\label{eq:osdef}\ .
\end{equation}
We now show that these matrix elements (and related ones) are
reproduced remarkably accurately by assuming the factorization
hypothesis and vacuum saturation. For each of these operators ($O_j$)
we define the ratio $R_j(t_1,t_2)$ analogously to eq.(\ref{eq:rjdef})
as follows:
\begin{equation}
R_j(t_1,t_2)\equiv\frac 38 \frac{K^{SS}_j(t_1,t_2)}{C^{LS}(-t_1)
C^{LS}(t_2)}\ ,
\label{eq:bbarrjdef}\end{equation}
where the correlation functions $C$ and $K_j$ are defined in
eqs.(\ref{eq:cdef}) and (\ref{eq:kdef}), except that in the $K_j$ the
operators $O_j$ are now $\Delta B=2$ operators, and the interpolating
operators destroy a $B$-meson and create a $\bar B$-meson.

We start by explaining explicitly what we mean by factorization
(combined with vacuum saturation). The evaluation of the B-parameter
requires the computation of three-point functions $K_j(t_1,t_2)$, which
are of the form~\footnote{The extension of this discussion to include
smeared interpolating operators is completely straightforward.}:
\begin{equation}
\langle\,0\,| \bar q(y)\gamma^5b(y)\ \bar b(0)\Gamma q(0)\,\bar
b(0)\tilde\Gamma q(0)\ \bar q(x)\gamma^5b(x)\,|\,0\,\rangle\ .
\label{eq:k3}\end{equation}
There are four Wick contractions which contribute to the correlation
function,  and it is convenient to track these by introducing a
fictitious quantum number, labelled by an integer suffix on each field,
so that the correlation function is written in the form
\begin{equation}
\langle\,0\,| \bar q_2(y)\gamma^5 b_1(y)\ O_{\Gamma,\tilde\Gamma}(0)\  
\bar q_4(x)\gamma^5 b_3(x)\,|\,0\,\rangle\ ,
\end{equation}
where 
\begin{equation}
O_{\Gamma\tilde\Gamma}\equiv
\bar b_1\Gamma q_2\ \bar b_3\tilde\Gamma q_4\ +
\bar b_3\Gamma q_4\ \bar b_1\tilde\Gamma q_2\ -
\bar b_1\Gamma q_4\ \bar b_3\tilde\Gamma q_2\ -
\bar b_3\Gamma q_2\ \bar b_1\tilde\Gamma q_4\ ,
\label{eq:ogg}\end{equation}
and $\Gamma,\tilde\Gamma$ are Dirac matrices. The contraction of
spinor ($\alpha,\beta$) and colour ($a$) indices in each bilinear in
eq.~(\ref{eq:ogg})  is implicit (e.g. $\bar b_1\Gamma q_2\,=\,
\bar b_{1,\,\alpha}^a\Gamma_{\alpha,\beta} q_{2,\,\beta}^a$).
Only contractions between fields with the same
suffix are allowed, and the four terms in eq.~(\ref{eq:ogg})
correspond to the four original Wick contractions. For all the
operators of interest, by using colour and spin Fierz identities, it
is possible to rewrite each of the four terms into sums of operators
of the form $(\bar b_1\Gamma^\prime q_2)\, (\bar
b_3\tilde\Gamma^\prime q_4)$ and $(\bar b_1\Gamma^\prime T^a q_2)
\,(\bar b_3\tilde\Gamma^\prime T^a q_4)$, for some $\gamma$-matrices
$\Gamma^\prime$ and $\tilde\Gamma^\prime$. In the factorization and
vacuum saturation hypothesis
\begin{equation}
\langle\, \bar B\,| (\bar b_1\Gamma^\prime T^a q_2)
\,(\bar b_3\tilde\Gamma^\prime T^a q_4)\,|\, B\,\rangle \simeq 0\ ,
\end{equation}
and
\begin{equation}
\langle\, \bar B\,| (\bar b_1\Gamma^\prime q_2)
\,(\bar b_3\tilde\Gamma^\prime q_4)\,|\, B\,\rangle \simeq
\langle\, \bar B\,| (\bar b_1\Gamma^\prime q_2)\,|\,0\rangle
\langle\,0\,|\,(\bar b_3\tilde\Gamma^\prime q_4)\,|\, B\,\rangle\ .
\label{eq:facts}\end{equation}
Lorentz invariance implies that each of the matrix
elements on the right-hand side of eq.(\ref{eq:facts}) vanishes or is
proportional to the leptonic decay constant $f_B$. 

We now consider each of the operators in turn:
\begin{itemize}
\item $O_L$ and $O_R$: The factor of 3/8 in eq.(\ref{eq:bbarrjdef})
was chosen so that the factorization hypothesis gives $R_L=R_R=1$.
Numerically we find:
\begin{equation}
\frac{R_L(3,3) + R_R(3,3)}{2} = 0.95\pm0.03\ .
\label{eq:rlresult}\end{equation}
\item $O_S$: Lorentz invariance implies that the matrix element
$\langle\,B\,|\bar b\sigma^{\mu\nu}(1-\gamma^5)q\,|\,0\,\rangle$
vanishes. Using this fact we deduce that factorization implies that
$R_S\simeq 5/8$. The numerical result for $R_S$ is 0.60(3), in very 
good agreement with the estimate based on factorization.
\item $O_N$: Finally factorization implies that $R_N\simeq 1$, in
good agreement with the numerical value 0.97(4). 
\end{itemize}

In order to illustrate further that the numerical results quoted above
are in agreement with expectations based on factorization and vacuum
saturation we present separately in fig.~\ref{fig:bbbar} the contributions
to each of the ratios from the two independent contractions:
\begin{equation}
R^a=\langle\,\bar B\,| (\bar b_1\Gamma^\prime q_2)
\,(\bar b_3\tilde\Gamma^\prime q_4)\,|\, B\,\rangle \ 
\end{equation}
and
\begin{equation}
R^b=\langle\, \bar B\,| (\bar b_1\Gamma^\prime q_4)
\,(\bar b_3\tilde\Gamma^\prime q_2)\,|\, B\,\rangle \ ,
\end{equation}
as well as the total combination
\begin{equation}
R=\frac{3}{8}(2 R^a - 2 R^b)
\end{equation}
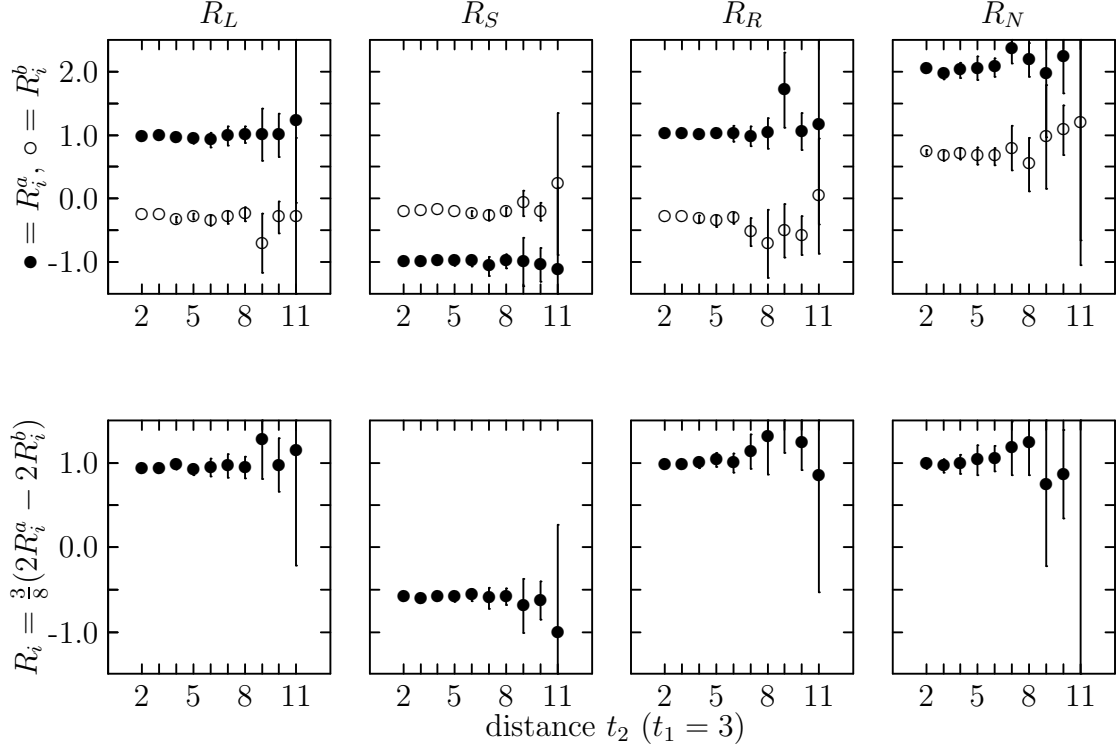
\begin{figure}[t]
\input{mydraw2.tex}
\caption{Results for the $R_j^a$, $R_j^b$ and $R_j$ ($j=L,S,R,N$)
as a function of $t_2$
for $t_1=3$. The value of the quark mass corresponds to $\kappa
=0.14226$.\label{fig:bbbar}}
\end{figure}
We see that not only are the values of the $R$'s quoted above in
agreement with expectations from factorization, but the values of each
of the $R^a$ and $R^b$ are also those we would expect on this
assumption: 
\begin{alignat}{5}
R_L^a& \simeq 1 &\qquad
R_L^b& \simeq -\frac{1}{3}R_L^a &\qquad;\qquad &
R_S^a& \simeq -1 &\qquad
R_S^b& \simeq \frac{1}{6}R_S^a\ \ \\ 
R_R^a& \simeq 1 &\qquad
R_R^b& \simeq -\frac{1}{3}R_L^a &\qquad;\qquad&
R_N^a& \simeq 2 &\qquad
R_N^b& \simeq \frac{1}{3}R_N^a\ .  
\end{alignat}

In this subsection we have demonstrated that the surprising precision of
predictions for matrix elements obtained using the factorization
hypothesis in inclusive decays, which was discussed in
section~\ref{sec:computation}, also applies to existing results for
$B$-$\bar B$ mixing. 

\subsection{Matching}
\label{subsec:matchingbb}

In this subsection we discuss very briefly the implications of the error
in the published value of the perturbative matching factors in $B$-$\bar
B$ mixing (see appendix B). For example, we take method (a) of
ref.~\cite{ewing}, in which each of the ratios $R_i$ is fitted separately,
and find
\begin{alignat}{2}
B_b(m_b) = 0.66(2)\ , & \qquad B_b = \alpha_s^{-6/23}B_b(m_b)=0.96(3)\ ,
\label{eq:bbnew}\end{alignat}
where $B_b(m_b)$ and $B_b$ are the $B$-parameter in the
$\overline{\mathrm MS}$ scheme at scale $m_b$ and the renormalization
group invariant $B$-parameter respectively. For the comparison we
only quote the statistical error. In order to see the effect of the error
in the matching coefficient, the result in eq.~(\ref{eq:bbnew}) should be 
compared with that obtained in ref.~\cite{ewing} using an identical procedure
(except for the values of the matching coefficients):
\begin{alignat}{2}
B_b(m_b) = 0.69(2)\ , &\qquad B_b =
\alpha_s^{-6/23}B_b(m_b)=1.02(3)\ .
\label{eq:bbold}\end{alignat} 

Gim\'enez and Martinelli had pointed out that the authors of
ref.~\cite{ewing} had used matching coefficient which did not include
the contributions to the $O(a^2)$ terms in the lattice operators which
they had used~\cite{gm}. As explained in ref.~\cite{gm}, this leads to
a negligible correction to the $B$-parameter. The differences in the
values in eqs.~(\ref{eq:bbnew}) and (\ref{eq:bbold}) is therefore
largely due to the error discussed in appendix B.

\section{Conclusions}
\label{sec:concs}

In this paper we have evaluated the matrix elements which contain the
non-perturbative QCD effects in the spectator contributions to inclusive
decays of $B$-mesons. For the \msbar\ scheme our principal results are
contained in eqs.~(\ref{eq:bmsbar}) and (\ref{eq:epsmsbar}). The raw
lattice results from which the matrix elements in any renormalization
scheme can be determined are given in
eqs.(\ref{eq:r12res})--(\ref{eq:r78res}).

The results for the matrix elements are very close to those which would
be expected on the basis of factorization and the vacuum saturation
hypothesis (particularly for the matrix elements of the bare lattice
operators). We find the extent to which this hypothesis is satisfied to
be surprising, but point out in subsec.~\ref{subsec:bbarfact} that this
was also the case for the $\Delta B=2$ operators which contribute to
$B$-$\bar B$ mixing, whose matrix elements have been computed by several
groups.

The calculation described in this paper is the first evaluation, using
lattice simulations, of the matrix elements of the operators in
eqs.(\ref{eq:ovadef})--(\ref{eq:tspdef}) between $B$-meson states.
The errors in the results presented in
eqs.(\ref{eq:r12res})--(\ref{eq:r78res}) are reasonably small, and
until unquenched computations become possible, it is not
phenomenologically necessary to invest a large effort to reduce the
statistical errors in this quenched calculation. It is desirable,
however, to establish, beyond any doubt, that the ground state meson
has been isolated and to confirm explicitly the validity of the
arguments presented in sec.~\ref{sec:computation} that this is indeed
the case. For this a similar calculation on a larger statistical
sample will be required. It would also be very useful to evaluate the
one-loop contributions to the Wilson coefficient functions in the OPE
for inclusive decay rates. It is probable that these corrections
currently represent the largest uncertainty in the inclusive rates.

A similar study of the matrix elements of the operators
eqs.(\ref{eq:r12res})--(\ref{eq:r78res}) between $\Lambda_b$ states is
in progress. This is particularly important in view of the discrepancy
between the experimental measurement in eq.~(\ref{eq:lbexp}) and the
theoretical prediction in eq.~(\ref{eq:lbth}) for the ratio
$\tau(\Lambda_b)/\tau(B_d)$. This calculation will help determine
whether the discrepancy is due to baryonic matrix elements being
larger than expected from quark models, or whether local quark-hadron
duality, assumed in the phenomenology, is not valid. Technically the
baryonic matrix elements of four-quark operators are more complicated
to evaluate than mesonic ones, since it is not sufficient to generate
the set of light-quark propagators from an arbitrary lattice point to
the origin. The results will be presented in a following publication.

\section*{Appendix A - Evaluation of the $D_{ij}$}

In this appendix we discuss the evaluation of the coefficients
$\{D_{ij}\}$ of eq.~(\ref{eq:matching2}) for a generic four-quark
operator $O_i$. This requires the evaluation of one loop corrections to
the matrix elements of the operators $\{O_i\}$ in both the continuum and
lattice schemes:
\begin{equation}
D_{ij}=C_{ij}^C-C_{ij}^L\ ,
\label{eq:dij}\end{equation}
where the superscripts $C$ and $L$ refer to the continuum
renormalization scheme ($\overline{MS}$) 
and the lattice regularization
respectively. We evaluate the matrix elements between four-quark
states at zero momentum, regulating the infrared divergences by giving
the gluon a small mass $\lambda$. The coefficients $\{D_{ij}\}$ do not
depend on the infra-red regulator, although each of
$C_{ij}^C$ and $C_{ij}^L$ are separately infrared
divergent.

We distinguish five categories of one-loop corrections to the generic
operator
\begin{equation}
O_i\equiv \bar b\Gamma^A_i q\,\bar q\tilde\Gamma^A_i b\ ,
\label{eq:oidef}\end{equation}
where $A$ represents both colour and spinor indices:
\begin{enumerate}
\item[1.] Corrections to the external lines. These are proportional to
$C_F$, the eigenvalue of the quadratic Casimir operator in the
fundamental representation ($C_F=4/3$ for the $SU(3)$ colour group), and
are independent of $\Gamma_i$ and $\tilde\Gamma_i$. 
\item[2.] Vertex corrections to each of the quark bilinears.
\item[3.] Corrections in which the gluon couples to both heavy-quark
propagators.
\item[4.] Corrections in which the gluon couples to one light-quark
propagator and one heavy-quark propagator (at the other vertex).
\item[5.] Corrections in which the gluon couples to both light-quark
propagators.
\end{enumerate}

In table~\ref{tab:cijcont} we present the values of the contributions
to the $C_{ij}^C$ in the $\overline{MS}$ scheme. For the
operators $O_1$ and $O_3$, $\Omega_1 = {-}\frac{1}{2}$ and
$\Omega_2=0$, whereas for $O_2$ and $O_4$ $\Omega_1 = 1$ and
$\Omega_2=0$. For a general choice of $O_i$, $\Omega_{1,2}$ depend on
the precise definition of the operator $\bar b(\Gamma^A_i
\sigma^{\mu\nu})q\,\bar q(\sigma^{\nu\mu}\tilde\Gamma^A_i)b$ and on
the choice of basis for the $\gamma$-matrices in D-dimensions.

\begin{table}[t]
\begin{center}
\begin{tabular}{|c|c|c|}\hline
(category)&$C_{ij}^C$ & $O_j^L$\rule{0pt}{13pt}\\ \hline
(1) & ${-}\log(\lambda^2a^2) + \frac{1}{2}$ &
$C_F\,\bar b\Gamma^A_i q\, \bar q\tilde\Gamma^A_i b$\rule[-8pt]{0pt}{22pt}\\
\hline
(2) & ${-}\log(\lambda^2a^2) + 1$ & $\bar b(T^a \Gamma^A_i T^a) q\, \bar
q\tilde\Gamma^A_i b +\bar b \Gamma^A_i q\, \bar q(T^a \tilde\Gamma^A_i
T^a) b $\rule{0pt}{14pt}\\ \hline
(3, odd) & $2\log(\lambda^2a^2)$ & $\bar b (T^a \Gamma^A_i) q\, \bar q
(\tilde\Gamma^A_i T^a) b$\rule{0pt}{14pt}\\ \hline
(4, odd) & $ \log(\lambda^2a^2) - 1$ & $\bar b (\Gamma^A_i T^a) q\, \bar
q(\tilde\Gamma^A_i T^a) b +\bar b (T^a \Gamma^A_i) q\, \bar q(T^a
\tilde\Gamma^A_i) b $\rule{0pt}{14pt}\\ \hline
(5, odd) & {-}$\log(\lambda^2a^2)+\Omega_1$ & $\bar b (\Gamma^A_i T^a) q\, \bar q
(T^a \tilde\Gamma^A_i) b$\rule{0pt}{14pt}\\ \hline
(5, odd) & {-}3$\log(\lambda^2a^2)+\Omega_2$ & $\frac{1}{12}\bar b
(\Gamma^A_i \sigma^{\mu\nu} T^a) q\, \bar q
(T^a \sigma^{\nu\mu}\tilde\Gamma^A_i) b$\rule[-8pt]{0pt}{22pt}\\ \hline
\end{tabular}
\caption{Values of the contributions to $C_{ij}^C$.
\label{tab:cijcont}}\end{center}
\end{table}

The coefficients $\{D_{ij}\}$ themselves are presented in
table~\ref{tab:dij}, where the $x_i$'s are defined as:
\begin{equation}
x_i\equiv c_i + c_i^I\ ,
\label{eq:dici}\end{equation}
and the numerical values of the $c_i$ and $c_i^I$ are tabulated in
table~\ref{tab:cis}. The contributions proportional to the $x_i$'s come
from the diagrams in the lattice regularization. The components
proportional to the $c_i$ would be the results if the Wilson formulation
of the quark action had been used, and those proportional to $c_i^I$ are
the additional contributions which result from the use of the
SW-improved action~\footnote{In the notation of ref.~\protect\cite{bp},
the parameters $\delta_v$ and $\delta_s$ are defined by
\[
\delta_v\equiv\frac 1{\pi ^2}\int_{-\pi }^{+\pi
}d^4k\,\left(\frac{\Delta_4(4-\Delta_1)-(\Delta_4-\Delta_5)}
{4\Delta _1\Delta _2^2}-3\frac{\theta (1-k^2)}{k^4}\right)
\ \ \ \ \textrm{and}
\]
\[
\delta_s\equiv\frac{r^2}{4\pi ^2}\int_{-\pi }^{+\pi }d^4k\frac{%
4\Delta _1(\Delta _4-\Delta _5)(2+r^2\Delta _1)-\Delta ^2_4
(2+r^2\Delta _1)}{4\Delta _1\Delta _2^2}\ .
\]}. 

\begin{table}[t]
\begin{center}
\begin{tabular}{|c|c|c|}\hline
j (Category)&$D_{ij}$ &
$O_j^L$\rule[-8pt]{0pt}{21pt}\\ \hline
1 (1,2) & $\frac{1}{2}{-}x_1$ & $C_F\,\bar b\Gamma^A_i q\, \bar
q\tilde\Gamma^A_i b$\rule[-8pt]{0pt}{22pt}\\ \hline
2 (2) & $1{-}x_2$ & $\bar b(T^a \Gamma^A_i T^a) q\, \bar
q\tilde\Gamma^A_i b +\bar b \Gamma^A_i q\, \bar q(T^a \tilde\Gamma^A_i
T^a) b $\rule{0pt}{14pt}\\ \hline
3 (2) & ${-}x_3$ & $\bar b(T^a \gamma^0 \Gamma^A_i \gamma^0 T^a) q\,
\bar q\tilde\Gamma^A_i b +\bar b \Gamma^A_i q\, \bar q(T^a \gamma^0
\tilde \Gamma^A_i\gamma^0 T^a) b $\rule{0pt}{14pt}\\ \hline
4 (3, odd) & ${-}x_4$ & $\bar b (T^a \Gamma^A_i) q\, \bar q
(\tilde\Gamma^A_i T^a) b$\rule{0pt}{14pt}\\ \hline
5 (4, odd) & ${-}1{-}x_5$ & $\bar b (\Gamma^A_i T^a) q\, \bar
q(\tilde\Gamma^A_i T^a) b +\bar b (T^a \Gamma^A_i) q\, \bar q(T^a
\tilde\Gamma^A_i) b $\rule{0pt}{14pt}\\ \hline
6 (4) & ${-}x_6$ & $\bar b (\Gamma^A_i\gamma^0 T^a) q\, \bar
q(\tilde\Gamma^A_i\gamma^0 T^a) b +\bar b (T^a \gamma^0\Gamma^A_i) q\,
\bar q(T^a \gamma^0\tilde\Gamma^A_i) b $\rule{0pt}{14pt}\\ \hline
7 (5, odd) & $\Omega_1{-}x_7$ & $\bar b (\Gamma^A_i T^a) q\, \bar q (T^a
\tilde\Gamma^A_i) b$\rule{0pt}{14pt}\\ \hline
8 (5, odd) & $\Omega_2{-}x_8$ & $\frac{1}{12}\bar b (\Gamma^A_i
\sigma^{\mu\nu} T^a) q\, \bar q (T^a \sigma^{\nu\mu}\tilde\Gamma^A_i)
b$\rule[-8pt]{0pt}{22pt}\\ \hline
9 (5) & ${-}x_9$ & $\frac{1}{4}\bar b (\Gamma^A_i\gamma^\mu T^a) q\,
\bar q (T^a\gamma^\mu\tilde\Gamma^A_i) b$\rule[-8pt]{0pt}{22pt}\\ \hline
\end{tabular}
\caption{Values of the contributions to $D_{ij}$.
\label{tab:dij}}\end{center} \end{table}

\begin{table}[ht]
\begin{center}
\begin{tabular}{|l|l|}
\hline
$c_1=f+e=17.89$ & $c_1^I=f^I-2(l+m)=-11.53$ \\ \hline
$c_2=d_1=5.46$ & $c_2^I=n=0.73$ \\ \hline
$c_3=d_2=-7.22$ & $c_3^I=h-2d^I-q=0.33$ \\ \hline
$c_4=-c=-4.53$ & $c_4^I=0$ \\ \hline
$c_5=-d_1=-5.46$ & $c_5^I=-n=-0.73$ \\ \hline
$c_6=d_2=-7.22$ & $c_6^I=c_3^I=0.33$ \\ \hline
$c_7=-v-\delta_v=4.85$ & $c_7^I=\delta_s=0.27$ \\ \hline
$c_8=\delta_v=2.07$ & $c_8^I=-v^I+s-\delta_s+3\delta_L=10.46$ \\ \hline
$c_9=4w=-4.84$ & $c_9^I=4w^I-2l-2m+s+3\delta_R=-4.88$ \\ \hline
\end{tabular}
\caption{Values of the coefficients $c_i$ and
$c^I_i$. Their expressions are given in terms of the
variables defined in refs.~\protect\cite{bp} and \protect\cite{bp2}
and two new variables $\delta_v$, $\delta_s$. 
The numerical integrals have been recomputed. Some of them differ
slightly from the results of \cite{bp}.
\label{tab:cis}}
\end{center}
\end{table}

\section*{Appendix B - $\Delta B=2$ operators}

There are many parallels between the calculations of spectator effects
in inclusive decays, which is the main subject of this paper, and that
of the matrix elements of the $\Delta B=2$ operators which contribute,
for example, to the important process of $B^0$-$\bar B^0$ mixing.
Indeed, we have recomputed the matrix elements of the $\Delta B=2$
operators and compared the results to those in ref.~\cite{ewing} as a
check on our procedures and programs. The calculation of the matching
factors are also similar in the two cases, and we have  exploited this
fact as a check on our perturbative calculation. Since we disagree with
one of the terms in the results of ref.~\cite{bp}, we briefly discuss
the evaluation of the matching factors in this appendix.

We consider a generic $\Delta B=2$ operator of the form $\bar
b\Gamma_i^Aq \,\bar b\tilde\Gamma^A_iq$. The evaluation of the various
contributions to the $D_{ij}$'s for $\Delta B=2$ operators parallels
that of the operator $O_i$ for spectator effects in inclusive decays
defined in eq.~(\ref{eq:oidef}); specifically, as explicitly marked in
table~\ref{tab:dij}, the coefficients corresponding to $j=4,5,7$ and 8
change sign, whilst the remaining coefficients are the same. Thus from
table~\ref{tab:dij} for spectator effects we deduce that the $D_{ij}$'s
for $\Delta B=2$ operators are as given in table~\ref{tab:dijdb2}. The
$x_i$'s are defined in eq.~(\ref{eq:dici}) and tabulated in
table~\ref{tab:cis}.

\begin{table}
\begin{center}
\begin{tabular}{|c|c|c|}
\hline
$j$ (category) & $D_{ij}^{\Delta B=2}$ & $O_j^L$ \\ \hline
$1$ (1,2) & $\frac 12-x_1$ & $C_F\,\overline{b}(\Gamma _i^A)q\,\overline{b}
(\widetilde{\Gamma }_i^A)q$ \\ \hline
$2$ (2) & $1-x_2$ & $\overline{b}(T^a\Gamma _i^AT^a)q\,\overline{b}(
\widetilde{\Gamma }_i^A)q+\overline{b}(\Gamma _i^A)q\,\overline{b}(T^a
\widetilde{\Gamma }_i^AT^a)q$ \\ \hline
$3$ (2) & $-x_3$ & $\overline{b}(T^a\gamma ^0\Gamma _i^A\gamma
^0T^a)q\,\overline{b}(\widetilde{\Gamma }_i^A)q+\overline{b}(\Gamma _i^A)q
\,\overline{b}(T^a\gamma ^0\widetilde{\Gamma }_i^A\gamma ^0T^a)q$ \\ \hline
$4$ (3, odd) & $x_4$ & $\overline{b}(T^a\Gamma _i^A)q\,\overline{b}
(T^a\widetilde{\Gamma }_i^A)q$ \\ \hline
$5$ (4, odd) & $1+x_5$ & $\overline{b}(\Gamma _i^AT^a)q\,\overline{b}
(T^a\widetilde{\Gamma }_i^A)q+\overline{b}(T^a\Gamma _i^A)q\,\overline{b}(
\widetilde{\Gamma }_i^AT^a)q$ \\ \hline
$6$ (4) & $-x_6$ & $\overline{b}(\Gamma _i^A\gamma ^0T^a)q\,\overline{b
}(T^a\gamma ^0\widetilde{\Gamma }_i^A)q+\overline{b}(T^a\gamma ^0\Gamma
_i^A)q\,\overline{b}(\widetilde{\Gamma }_i^A\gamma ^0T^a)q$ \\ \hline
$7$ (5, odd) & $-\Omega _1+x_7$ & $\overline{b}(\Gamma _i^AT^a)q\,
\overline{b}(\widetilde{\Gamma }_i^AT^a)q$ \\ \hline
$8$ (5, odd) & $-\Omega _2+x_8$ & $\frac 1{12}\overline{b}(\Gamma
_i^A\sigma ^{\mu \nu }T^a)q\,\overline{b}(\widetilde{\Gamma }_i^A\sigma ^{\mu
\nu }T^a)q$ \\ \hline
$9$ (5) & $-x_9$ & $\frac 14\overline{b}(\Gamma _i^A\gamma ^\mu T^a)q
\,\overline{b}(\widetilde{\Gamma }_i^A\gamma ^\mu T^a)q$ \\ \hline
\end{tabular}
\caption{Values of the contributions to the $D_{ij}$'s for $\Delta B=2$ operators.
\label{tab:dijdb2}}
\end{center}
\end{table}

We are particularly interested in the operator whose matrix elements
contains the non-perturbative QCD effects for $B^0$-$\bar B^0$ mixing:
\begin{equation}
\bar b\Gamma^A_i q\,\bar b\widetilde\Gamma^Aq \equiv O_L
= \bar b\gamma^\rho q_L\,\bar b\gamma_\rho q_L\ ,
\label{eq:oldef}\end{equation}
where the label $L$ denotes ``left". In this case the different
operators $O_j$ of table~\ref{tab:dijdb2} reduce to 3 independent
ones: 
\begin{eqnarray}
O_R & = & \bar b\gamma^\rho q_R\,\bar b\gamma_\rho q_R\label{eq:ordef}\\
O_N & = & 2\bar b q_L\,\bar b q_R +  2\bar b q_R\,\bar b q_L +
\bar b \gamma^\rho q_L\,\bar b \gamma_\rho q_R +
\bar b \gamma^\rho q_R\,\bar b \gamma_\rho q_L\ ,\label{eq:ondef}
\end{eqnarray}
as well as $O_L$ itself. Step i) of the matching
(see section~\ref{sec:matching}) is now given by the relation:
\begin{equation}
O_L^C =(1+\frac{\alpha _s}{4\pi}D_L)O_L^L+\frac{
\alpha _s}{4\pi}D_N O_N^L+\frac{\alpha _s}{4\pi}
D_R O_R^L\label{eq:resbb}
\end{equation}
where
\begin{alignat}{4}
D_L = -22.4\ ,\ &\   
D_N = -13.8\ \ &\ \textrm{and}\    
D_R = -3.2\ \label{eq:dsres}\ .  
\end{alignat}
For the continuum renormalization scheme we have used $\overline{MS}$.
In $4-2\epsilon$ dimensions
\begin{equation}
\bar b\gamma^\rho\sigma^{\mu\nu}q_L\,
\bar b\gamma_\rho\sigma_{\mu\nu}q_L = (12 - 2\epsilon)
\bar b\gamma^\rho q_L\,\bar b\gamma_\rho q_L\ ,
\end{equation} 
from which we derive that $\Omega_1+\Omega_2=5$ in this scheme. In
considering the crossing relations between inclusive decays and
mixing, we note that in the latter process, one of the $\bar b$ fields
destroys a quark (so that $\bar b\gamma^0=\bar b$) and the other
creates an antiquark  (so that $\bar b\gamma^0=-\bar b$).

The results for $D_L$ and $D_N$ in eq.~(\ref{eq:dsres}) agree with those
in the literature~\cite{bp2}, whereas that for $D_R$ does not (a similar
conclusion was reached independently by Gim\'enez~\cite{gimenez}). In
ref.~\cite{bp2} the quoted result is $D_R=-5.4$~\footnote{We believe
that the reason is that in eq.~(B.16) of ref.\cite{bp} there should be a
correction:
\begin{equation*}
(...)+\frac{g^2}{16\pi^2}\frac 43\,(\omega + \omega^I)
\to 
(...)-\frac{g^2}{16\pi^2}\frac 43\,(\omega + \omega^I)
\,O^{latt}_R\ ,
\end{equation*}
eq.~(B.26) should be replaced by
\begin{equation*}
D_R^I=\frac 13\,[s+4\omega^I-2(l+m)]
\end{equation*}
and that in table 3
\begin{equation*}
D_R^I=-0.38\ \ \textrm{for}\ r=1\ .
\end{equation*}
There errors are carried forward into successive papers \cite{bp2},
\cite{ewing} and \cite{gm}. 
Moreover in the same paper, in the definition of $v^I$ (eq.~(B.17)), 
the term $\Delta_2$ at the denominator should be replaced by $\Delta_2^2$. 
}.

\section*{Acknowledgements}

We thank Vicente Gimen\'ez and Juan Reyes for detailed correspondence
concerning the matching factors and for pointing out some errors in
the earlier version of this paper. We also thank Carlotta Pittori,
Giulia De Divitiis, Luigi Del Debbio and Jonathan Flynn for many helpful
discussion. We gratefully acknowledge Hartmut Wittig and other 
colleagues from the UKQCD collaboration for enabling us to use the
gauge configurations and quark propagators.

This work was supported by PPARC grants GR/L29927 and  GR/L56329, and
EPSRC grant GR/K41663.

\end{document}

%% file: mydraw.tex
\begin{center}                                  
                \makebox[0pt]{\begin{texdraw}
                \normalsize
                \ifx\pathDEFINED\relax\else\let\pathDEFINED\relax
                 \def\QtGfr{\ifx (\TGre \let\YhetT\cpath\else\let\YhetT\relax\fi\YhetT}
                 \def\path (#1 #2){\move (#1 #2)\futurelet\TGre\QtGfr}
                 \def\cpath (#1 #2){\lvec (#1 #2)\futurelet\TGre\QtGfr}
                \fi
                \drawdim pt
                \setunitscale 0.24
                \linewd 3
\move(950 625) \textref h:C v:C \htext{}
\move(25 910) \textref h:C v:C \vtext{$R_1,R_2, -R_5,-R_6$}
\linewd 3
\path (150 710)(150 1110)
\textref h:C v:C
\path (150 710)(165 710)
\path (150 910)(165 910)
\path (150 1110)(165 1110)
\textref h:R v:C
\move (135 710) \htext{0.5}
\move (135 910) \htext{1.0}
\move (135 1110) \htext{1.5}
\textref h:C v:C
\move(325 625) \textref h:C v:C \htext{}
\move(25 910) \textref h:C v:C \vtext{}
\linewd 3
\path (150 710)(500 710)
\path (150 1110)(500 1110)
\path (150 710)(150 1110)
\path (500 710)(500 1110)
\path (203 710)(203 725)
\path (203 1110)(203 1095)
\path (230 710)(230 725)
\path (230 1110)(230 1095)
\path (257 710)(257 725)
\path (257 1110)(257 1095)
\path (284 710)(284 725)
\path (284 1110)(284 1095)
\path (311 710)(311 725)
\path (311 1110)(311 1095)
\path (338 710)(338 725)
\path (338 1110)(338 1095)
\path (365 710)(365 725)
\path (365 1110)(365 1095)
\path (392 710)(392 725)
\path (392 1110)(392 1095)
\path (419 710)(419 725)
\path (419 1110)(419 1095)
\path (446 710)(446 725)
\path (446 1110)(446 1095)
\textref h:C v:C
\move (203 675) \htext{2}
\move (284 675) \htext{5}
\move (365 675) \htext{8}
\move (446 675) \htext{11}
\path (150 710)(165 710)
\path (500 710)(485 710)
\path (150 910)(165 910)
\path (500 910)(485 910)
\path (150 1110)(165 1110)
\path (500 1110)(485 1110)
\textref h:R v:C
\textref h:C v:C
\move(325 1150) \textref h:C v:C \htext{$\overline{b}_L\gamma^\mu q_L \overline{q}_L \gamma^\mu b_L$}
\textref h:C v:C
\move (203 917) \htext{$\bullet$}
\textref h:C v:C
\move (230 920) \htext{$\bullet$}
\textref h:C v:C
\path (257 891)(257 924)
\path (257 891)(257 891)
\path (257 924)(257 924)
\move (257 907) \htext{$\bullet$}
\textref h:C v:C
\path (284 861)(284 912)
\path (284 861)(284 861)
\path (284 912)(284 912)
\move (284 886) \htext{$\bullet$}
\textref h:C v:C
\path (311 876)(311 946)
\path (311 876)(311 876)
\path (311 946)(311 946)
\move (311 911) \htext{$\bullet$}
\textref h:C v:C
\path (338 916)(338 1003)
\path (338 916)(338 916)
\path (338 1003)(338 1003)
\move (338 960) \htext{$\bullet$}
\textref h:C v:C
\path (365 886)(365 1003)
\path (365 886)(365 886)
\path (365 1003)(365 1003)
\move (365 944) \htext{$\bullet$}
\textref h:C v:C
\path (392 792)(392 1110)
\path (392 792)(392 792)
\path (392 1110)(392 1110)
\move (392 1033) \htext{$\bullet$}
\textref h:C v:C
\path (419 944)(419 1110)
\path (419 944)(419 944)
\path (419 1110)(419 1110)
\move (419 1055) \htext{$\bullet$}
\textref h:C v:C
\path (446 710)(446 1110)
\path (446 710)(446 710)
\path (446 1110)(446 1110)
\move(737 625) \textref h:C v:C \htext{}
\move(437 910) \textref h:C v:C \vtext{}
\linewd 3
\path (562 710)(912 710)
\path (562 1110)(912 1110)
\path (562 710)(562 1110)
\path (912 710)(912 1110)
\path (615 710)(615 725)
\path (615 1110)(615 1095)
\path (642 710)(642 725)
\path (642 1110)(642 1095)
\path (669 710)(669 725)
\path (669 1110)(669 1095)
\path (696 710)(696 725)
\path (696 1110)(696 1095)
\path (723 710)(723 725)
\path (723 1110)(723 1095)
\path (750 710)(750 725)
\path (750 1110)(750 1095)
\path (777 710)(777 725)
\path (777 1110)(777 1095)
\path (804 710)(804 725)
\path (804 1110)(804 1095)
\path (831 710)(831 725)
\path (831 1110)(831 1095)
\path (858 710)(858 725)
\path (858 1110)(858 1095)
\textref h:C v:C
\move (615 675) \htext{2}
\move (696 675) \htext{5}
\move (777 675) \htext{8}
\move (858 675) \htext{11}
\path (562 710)(577 710)
\path (912 710)(897 710)
\path (562 910)(577 910)
\path (912 910)(897 910)
\path (562 1110)(577 1110)
\path (912 1110)(897 1110)
\textref h:R v:C
\textref h:C v:C
\move(737 1150) \textref h:C v:C \htext{$\overline{b}_R q_L \overline{q}_L b_R$}
\textref h:C v:C
\move (615 914) \htext{$\bullet$}
\textref h:C v:C
\move (642 910) \htext{$\bullet$}
\textref h:C v:C
\move (669 911) \htext{$\bullet$}
\textref h:C v:C
\path (696 895)(696 933)
\path (696 895)(696 895)
\path (696 933)(696 933)
\move (696 914) \htext{$\bullet$}
\textref h:C v:C
\path (723 892)(723 945)
\path (723 892)(723 892)
\path (723 945)(723 945)
\move (723 918) \htext{$\bullet$}
\textref h:C v:C
\path (750 934)(750 1009)
\path (750 934)(750 934)
\path (750 1009)(750 1009)
\move (750 972) \htext{$\bullet$}
\textref h:C v:C
\path (777 879)(777 965)
\path (777 879)(777 879)
\path (777 965)(777 965)
\move (777 922) \htext{$\bullet$}
\textref h:C v:C
\path (804 820)(804 1088)
\path (804 820)(804 820)
\path (804 1088)(804 1088)
\move (804 954) \htext{$\bullet$}
\textref h:C v:C
\path (831 846)(831 1052)
\path (831 846)(831 846)
\path (831 1052)(831 1052)
\move (831 949) \htext{$\bullet$}
\textref h:C v:C
\path (858 710)(858 1110)
\path (858 710)(858 710)
\path (858 1110)(858 1110)
\move (858 1050) \htext{$\bullet$}
\move(1149 625) \textref h:C v:C \htext{}
\move(849 910) \textref h:C v:C \vtext{}
\linewd 3
\path (974 710)(1324 710)
\path (974 1110)(1324 1110)
\path (974 710)(974 1110)
\path (1324 710)(1324 1110)
\path (1027 710)(1027 725)
\path (1027 1110)(1027 1095)
\path (1054 710)(1054 725)
\path (1054 1110)(1054 1095)
\path (1081 710)(1081 725)
\path (1081 1110)(1081 1095)
\path (1108 710)(1108 725)
\path (1108 1110)(1108 1095)
\path (1135 710)(1135 725)
\path (1135 1110)(1135 1095)
\path (1162 710)(1162 725)
\path (1162 1110)(1162 1095)
\path (1189 710)(1189 725)
\path (1189 1110)(1189 1095)
\path (1216 710)(1216 725)
\path (1216 1110)(1216 1095)
\path (1243 710)(1243 725)
\path (1243 1110)(1243 1095)
\path (1270 710)(1270 725)
\path (1270 1110)(1270 1095)
\textref h:C v:C
\move (1027 675) \htext{2}
\move (1108 675) \htext{5}
\move (1189 675) \htext{8}
\move (1270 675) \htext{11}
\path (974 710)(989 710)
\path (1324 710)(1309 710)
\path (974 910)(989 910)
\path (1324 910)(1309 910)
\path (974 1110)(989 1110)
\path (1324 1110)(1309 1110)
\textref h:R v:C
\textref h:C v:C
\move(1149 1150) \textref h:C v:C \htext{$-\overline{b}_R\gamma^\mu q_R \overline{q}_L \gamma^\mu b_L$}
\textref h:C v:C
\move (1027 902) \htext{$\bullet$}
\textref h:C v:C
\path (1054 889)(1054 921)
\path (1054 889)(1054 889)
\path (1054 921)(1054 921)
\move (1054 905) \htext{$\bullet$}
\textref h:C v:C
\path (1081 872)(1081 908)
\path (1081 872)(1081 872)
\path (1081 908)(1081 908)
\move (1081 890) \htext{$\bullet$}
\textref h:C v:C
\path (1108 862)(1108 923)
\path (1108 862)(1108 862)
\path (1108 923)(1108 923)
\move (1108 893) \htext{$\bullet$}
\textref h:C v:C
\path (1135 872)(1135 970)
\path (1135 872)(1135 872)
\path (1135 970)(1135 970)
\move (1135 921) \htext{$\bullet$}
\textref h:C v:C
\path (1162 895)(1162 1013)
\path (1162 895)(1162 895)
\path (1162 1013)(1162 1013)
\move (1162 954) \htext{$\bullet$}
\textref h:C v:C
\path (1189 843)(1189 963)
\path (1189 843)(1189 843)
\path (1189 963)(1189 963)
\move (1189 903) \htext{$\bullet$}
\textref h:C v:C
\path (1216 710)(1216 1062)
\path (1216 710)(1216 710)
\path (1216 1062)(1216 1062)
\move (1216 857) \htext{$\bullet$}
\textref h:C v:C
\path (1243 710)(1243 910)
\path (1243 710)(1243 710)
\path (1243 910)(1243 910)
\move (1243 791) \htext{$\bullet$}
\textref h:C v:C
\path (1270 710)(1270 1110)
\path (1270 710)(1270 710)
\path (1270 1110)(1270 1110)
\move (1270 876) \htext{$\bullet$}
\move(1561 625) \textref h:C v:C \htext{}
\move(1261 910) \textref h:C v:C \vtext{}
\linewd 3
\path (1386 710)(1736 710)
\path (1386 1110)(1736 1110)
\path (1386 710)(1386 1110)
\path (1736 710)(1736 1110)
\path (1439 710)(1439 725)
\path (1439 1110)(1439 1095)
\path (1466 710)(1466 725)
\path (1466 1110)(1466 1095)
\path (1493 710)(1493 725)
\path (1493 1110)(1493 1095)
\path (1520 710)(1520 725)
\path (1520 1110)(1520 1095)
\path (1547 710)(1547 725)
\path (1547 1110)(1547 1095)
\path (1574 710)(1574 725)
\path (1574 1110)(1574 1095)
\path (1601 710)(1601 725)
\path (1601 1110)(1601 1095)
\path (1628 710)(1628 725)
\path (1628 1110)(1628 1095)
\path (1655 710)(1655 725)
\path (1655 1110)(1655 1095)
\path (1682 710)(1682 725)
\path (1682 1110)(1682 1095)
\textref h:C v:C
\move (1439 675) \htext{2}
\move (1520 675) \htext{5}
\move (1601 675) \htext{8}
\move (1682 675) \htext{11}
\path (1386 710)(1401 710)
\path (1736 710)(1721 710)
\path (1386 910)(1401 910)
\path (1736 910)(1721 910)
\path (1386 1110)(1401 1110)
\path (1736 1110)(1721 1110)
\textref h:R v:C
\textref h:C v:C
\move(1561 1150) \textref h:C v:C \htext{$-\overline{b}_R q_L \overline{q}_R b_L$}
\textref h:C v:C
\move (1439 900) \htext{$\bullet$}
\textref h:C v:C
\move (1466 905) \htext{$\bullet$}
\textref h:C v:C
\move (1493 895) \htext{$\bullet$}
\textref h:C v:C
\path (1520 880)(1520 930)
\path (1520 880)(1520 880)
\path (1520 930)(1520 930)
\move (1520 905) \htext{$\bullet$}
\textref h:C v:C
\path (1547 889)(1547 950)
\path (1547 889)(1547 889)
\path (1547 950)(1547 950)
\move (1547 920) \htext{$\bullet$}
\textref h:C v:C
\path (1574 897)(1574 997)
\path (1574 897)(1574 897)
\path (1574 997)(1574 997)
\move (1574 947) \htext{$\bullet$}
\textref h:C v:C
\path (1601 868)(1601 960)
\path (1601 868)(1601 868)
\path (1601 960)(1601 960)
\move (1601 914) \htext{$\bullet$}
\textref h:C v:C
\path (1628 725)(1628 1053)
\path (1628 725)(1628 725)
\path (1628 1053)(1628 1053)
\move (1628 889) \htext{$\bullet$}
\textref h:C v:C
\path (1655 711)(1655 959)
\path (1655 711)(1655 711)
\path (1655 959)(1655 959)
\move (1655 835) \htext{$\bullet$}
\textref h:C v:C
\path (1682 710)(1682 1110)
\path (1682 710)(1682 710)
\path (1682 1110)(1682 1110)
\move (1682 950) \htext{$\bullet$}
\move(950 25) \textref h:C v:C \htext{distance $t_2$ $(t_1=3)$}
\move(25 310) \textref h:C v:C \vtext{$R_{3,4,7,8}$}
\linewd 3
\path (150 110)(150 510)
\textref h:C v:C
\path (150 110)(165 110)
\path (150 310)(165 310)
\path (150 510)(165 510)
\textref h:R v:C
\move (135 110) \htext{-0.5}
\move (135 310) \htext{0.0}
\move (135 510) \htext{0.5}
\textref h:C v:C
\move(325 25) \textref h:C v:C \htext{}
\move(25 310) \textref h:C v:C \vtext{}
\linewd 3
\path (150 110)(500 110)
\path (150 510)(500 510)
\path (150 110)(150 510)
\path (500 110)(500 510)
\path (203 110)(203 125)
\path (203 510)(203 495)
\path (230 110)(230 125)
\path (230 510)(230 495)
\path (257 110)(257 125)
\path (257 510)(257 495)
\path (284 110)(284 125)
\path (284 510)(284 495)
\path (311 110)(311 125)
\path (311 510)(311 495)
\path (338 110)(338 125)
\path (338 510)(338 495)
\path (365 110)(365 125)
\path (365 510)(365 495)
\path (392 110)(392 125)
\path (392 510)(392 495)
\path (419 110)(419 125)
\path (419 510)(419 495)
\path (446 110)(446 125)
\path (446 510)(446 495)
\textref h:C v:C
\move (203 75) \htext{2}
\move (284 75) \htext{5}
\move (365 75) \htext{8}
\move (446 75) \htext{11}
\path (150 110)(165 110)
\path (500 110)(485 110)
\path (150 310)(165 310)
\path (500 310)(485 310)
\path (150 510)(165 510)
\path (500 510)(485 510)
\textref h:R v:C
\textref h:C v:C
\move(325 550) \textref h:C v:C \htext{$\overline{b}_L\gamma^\mu T^a q_L\overline{q}_L\gamma^\mu T^a b_L$}
\textref h:C v:C
\move (203 306) \htext{$\bullet$}
\textref h:C v:C
\move (230 310) \htext{$\bullet$}
\textref h:C v:C
\move (257 309) \htext{$\bullet$}
\textref h:C v:C
\move (284 305) \htext{$\bullet$}
\textref h:C v:C
\move (311 310) \htext{$\bullet$}
\textref h:C v:C
\path (338 264)(338 311)
\path (338 264)(338 264)
\path (338 311)(338 311)
\move (338 287) \htext{$\bullet$}
\textref h:C v:C
\path (365 281)(365 327)
\path (365 281)(365 281)
\path (365 327)(365 327)
\move (365 304) \htext{$\bullet$}
\textref h:C v:C
\path (392 280)(392 420)
\path (392 280)(392 280)
\path (392 420)(392 420)
\move (392 350) \htext{$\bullet$}
\textref h:C v:C
\path (419 196)(419 298)
\path (419 196)(419 196)
\path (419 298)(419 298)
\move (419 247) \htext{$\bullet$}
\textref h:C v:C
\path (446 110)(446 413)
\path (446 110)(446 110)
\path (446 413)(446 413)
\move (446 204) \htext{$\bullet$}
\move(737 25) \textref h:C v:C \htext{}
\move(437 310) \textref h:C v:C \vtext{}
\linewd 3
\path (562 110)(912 110)
\path (562 510)(912 510)
\path (562 110)(562 510)
\path (912 110)(912 510)
\path (615 110)(615 125)
\path (615 510)(615 495)
\path (642 110)(642 125)
\path (642 510)(642 495)
\path (669 110)(669 125)
\path (669 510)(669 495)
\path (696 110)(696 125)
\path (696 510)(696 495)
\path (723 110)(723 125)
\path (723 510)(723 495)
\path (750 110)(750 125)
\path (750 510)(750 495)
\path (777 110)(777 125)
\path (777 510)(777 495)
\path (804 110)(804 125)
\path (804 510)(804 495)
\path (831 110)(831 125)
\path (831 510)(831 495)
\path (858 110)(858 125)
\path (858 510)(858 495)
\textref h:C v:C
\move (615 75) \htext{2}
\move (696 75) \htext{5}
\move (777 75) \htext{8}
\move (858 75) \htext{11}
\path (562 110)(577 110)
\path (912 110)(897 110)
\path (562 310)(577 310)
\path (912 310)(897 310)
\path (562 510)(577 510)
\path (912 510)(897 510)
\textref h:R v:C
\textref h:C v:C
\move(737 550) \textref h:C v:C \htext{$\overline{b}_R T^a q_L \overline{q}_L T^a b_R$}
\textref h:C v:C
\move (615 310) \htext{$\bullet$}
\textref h:C v:C
\move (642 309) \htext{$\bullet$}
\textref h:C v:C
\move (669 309) \htext{$\bullet$}
\textref h:C v:C
\move (696 308) \htext{$\bullet$}
\textref h:C v:C
\move (723 305) \htext{$\bullet$}
\textref h:C v:C
\move (750 308) \htext{$\bullet$}
\textref h:C v:C
\path (777 278)(777 311)
\path (777 278)(777 278)
\path (777 311)(777 311)
\move (777 295) \htext{$\bullet$}
\textref h:C v:C
\path (804 288)(804 320)
\path (804 288)(804 288)
\path (804 320)(804 320)
\move (804 304) \htext{$\bullet$}
\textref h:C v:C
\move (831 296) \htext{$\bullet$}
\textref h:C v:C
\path (858 140)(858 348)
\path (858 140)(858 140)
\path (858 348)(858 348)
\move (858 244) \htext{$\bullet$}
\move(1149 25) \textref h:C v:C \htext{}
\move(849 310) \textref h:C v:C \vtext{}
\linewd 3
\path (974 110)(1324 110)
\path (974 510)(1324 510)
\path (974 110)(974 510)
\path (1324 110)(1324 510)
\path (1027 110)(1027 125)
\path (1027 510)(1027 495)
\path (1054 110)(1054 125)
\path (1054 510)(1054 495)
\path (1081 110)(1081 125)
\path (1081 510)(1081 495)
\path (1108 110)(1108 125)
\path (1108 510)(1108 495)
\path (1135 110)(1135 125)
\path (1135 510)(1135 495)
\path (1162 110)(1162 125)
\path (1162 510)(1162 495)
\path (1189 110)(1189 125)
\path (1189 510)(1189 495)
\path (1216 110)(1216 125)
\path (1216 510)(1216 495)
\path (1243 110)(1243 125)
\path (1243 510)(1243 495)
\path (1270 110)(1270 125)
\path (1270 510)(1270 495)
\textref h:C v:C
\move (1027 75) \htext{2}
\move (1108 75) \htext{5}
\move (1189 75) \htext{8}
\move (1270 75) \htext{11}
\path (974 110)(989 110)
\path (1324 110)(1309 110)
\path (974 310)(989 310)
\path (1324 310)(1309 310)
\path (974 510)(989 510)
\path (1324 510)(1309 510)
\textref h:R v:C
\textref h:C v:C
\move(1149 550) \textref h:C v:C \htext{$\overline{b}_R\gamma^\mu T^a q_R\overline{q}_L\gamma^\mu T^a b_L$}
\textref h:C v:C
\move (1027 295) \htext{$\bullet$}
\textref h:C v:C
\move (1054 294) \htext{$\bullet$}
\textref h:C v:C
\move (1081 307) \htext{$\bullet$}
\textref h:C v:C
\move (1108 314) \htext{$\bullet$}
\textref h:C v:C
\path (1135 289)(1135 328)
\path (1135 289)(1135 289)
\path (1135 328)(1135 328)
\move (1135 309) \htext{$\bullet$}
\textref h:C v:C
\path (1162 287)(1162 346)
\path (1162 287)(1162 287)
\path (1162 346)(1162 346)
\move (1162 317) \htext{$\bullet$}
\textref h:C v:C
\path (1189 298)(1189 358)
\path (1189 298)(1189 298)
\path (1189 358)(1189 358)
\move (1189 328) \htext{$\bullet$}
\textref h:C v:C
\path (1216 339)(1216 510)
\path (1216 339)(1216 339)
\path (1216 510)(1216 510)
\move (1216 452) \htext{$\bullet$}
\textref h:C v:C
\path (1243 331)(1243 489)
\path (1243 331)(1243 331)
\path (1243 489)(1243 489)
\move (1243 410) \htext{$\bullet$}
\textref h:C v:C
\path (1270 110)(1270 510)
\path (1270 110)(1270 110)
\path (1270 510)(1270 510)
\move (1270 296) \htext{$\bullet$}
\move(1561 25) \textref h:C v:C \htext{}
\move(1261 310) \textref h:C v:C \vtext{}
\linewd 3
\path (1386 110)(1736 110)
\path (1386 510)(1736 510)
\path (1386 110)(1386 510)
\path (1736 110)(1736 510)
\path (1439 110)(1439 125)
\path (1439 510)(1439 495)
\path (1466 110)(1466 125)
\path (1466 510)(1466 495)
\path (1493 110)(1493 125)
\path (1493 510)(1493 495)
\path (1520 110)(1520 125)
\path (1520 510)(1520 495)
\path (1547 110)(1547 125)
\path (1547 510)(1547 495)
\path (1574 110)(1574 125)
\path (1574 510)(1574 495)
\path (1601 110)(1601 125)
\path (1601 510)(1601 495)
\path (1628 110)(1628 125)
\path (1628 510)(1628 495)
\path (1655 110)(1655 125)
\path (1655 510)(1655 495)
\path (1682 110)(1682 125)
\path (1682 510)(1682 495)
\textref h:C v:C
\move (1439 75) \htext{2}
\move (1520 75) \htext{5}
\move (1601 75) \htext{8}
\move (1682 75) \htext{11}
\path (1386 110)(1401 110)
\path (1736 110)(1721 110)
\path (1386 310)(1401 310)
\path (1736 310)(1721 310)
\path (1386 510)(1401 510)
\path (1736 510)(1721 510)
\textref h:R v:C
\textref h:C v:C
\move(1561 550) \textref h:C v:C \htext{$\overline{b}_R T^a q_L \overline{q}_R T^a b_L$}
\textref h:C v:C
\move (1439 308) \htext{$\bullet$}
\textref h:C v:C
\move (1466 308) \htext{$\bullet$}
\textref h:C v:C
\move (1493 318) \htext{$\bullet$}
\textref h:C v:C
\move (1520 319) \htext{$\bullet$}
\textref h:C v:C
\move (1547 312) \htext{$\bullet$}
\textref h:C v:C
\move (1574 310) \htext{$\bullet$}
\textref h:C v:C
\move (1601 331) \htext{$\bullet$}
\textref h:C v:C
\path (1628 318)(1628 430)
\path (1628 318)(1628 318)
\path (1628 430)(1628 430)
\move (1628 374) \htext{$\bullet$}
\textref h:C v:C
\path (1655 324)(1655 404)
\path (1655 324)(1655 324)
\path (1655 404)(1655 404)
\move (1655 364) \htext{$\bullet$}
\textref h:C v:C
\path (1682 198)(1682 510)
\path (1682 198)(1682 198)
\path (1682 510)(1682 510)
\move (1682 427) \htext{$\bullet$}
\end{texdraw}}
\end{center}

%% file: mydraw2.tex
\begin{center}                                  
                \makebox[0pt]{\begin{texdraw}
                \normalsize
                \ifx\pathDEFINED\relax\else\let\pathDEFINED\relax
                 \def\QtGfr{\ifx (\TGre \let\YhetT\cpath\else\let\YhetT\relax\fi\YhetT}
                 \def\path (#1 #2){\move (#1 #2)\futurelet\TGre\QtGfr}
                 \def\cpath (#1 #2){\lvec (#1 #2)\futurelet\TGre\QtGfr}
                \fi
                \drawdim pt
                \setunitscale 0.24
                \linewd 3
\move(950 625) \textref h:C v:C \htext{}
\move(25 910) \textref h:C v:C \vtext{$\bullet=R_i^a$, $\circ=R_i^b$}
\linewd 3
\path (150 710)(150 1110)
\textref h:C v:C
\path (150 760)(165 760)
\path (150 810)(165 810)
\path (150 860)(165 860)
\path (150 910)(165 910)
\path (150 960)(165 960)
\path (150 1010)(165 1010)
\path (150 1060)(165 1060)
\textref h:R v:C
\move (135 760) \htext{-1.0}
\move (135 860) \htext{0.0}
\move (135 960) \htext{1.0}
\move (135 1060) \htext{2.0}
\textref h:C v:C
\move(325 625) \textref h:C v:C \htext{}
\move(25 910) \textref h:C v:C \vtext{}
\linewd 3
\path (150 710)(500 710)
\path (150 1110)(500 1110)
\path (150 710)(150 1110)
\path (500 710)(500 1110)
\path (203 710)(203 725)
\path (203 1110)(203 1095)
\path (230 710)(230 725)
\path (230 1110)(230 1095)
\path (257 710)(257 725)
\path (257 1110)(257 1095)
\path (284 710)(284 725)
\path (284 1110)(284 1095)
\path (311 710)(311 725)
\path (311 1110)(311 1095)
\path (338 710)(338 725)
\path (338 1110)(338 1095)
\path (365 710)(365 725)
\path (365 1110)(365 1095)
\path (392 710)(392 725)
\path (392 1110)(392 1095)
\path (419 710)(419 725)
\path (419 1110)(419 1095)
\path (446 710)(446 725)
\path (446 1110)(446 1095)
\textref h:C v:C
\move (203 675) \htext{2}
\move (284 675) \htext{5}
\move (365 675) \htext{8}
\move (446 675) \htext{11}
\path (150 760)(165 760)
\path (500 760)(485 760)
\path (150 810)(165 810)
\path (500 810)(485 810)
\path (150 860)(165 860)
\path (500 860)(485 860)
\path (150 910)(165 910)
\path (500 910)(485 910)
\path (150 960)(165 960)
\path (500 960)(485 960)
\path (150 1010)(165 1010)
\path (500 1010)(485 1010)
\path (150 1060)(165 1060)
\path (500 1060)(485 1060)
\textref h:R v:C
\textref h:C v:C
\move(325 1150) \textref h:C v:C \htext{$R_L$}
\textref h:C v:C
\move (203 958) \htext{$\bullet$}
\textref h:C v:C
\path (230 954)(230 963)
\path (230 954)(230 954)
\path (230 963)(230 963)
\move (230 959) \htext{$\bullet$}
\textref h:C v:C
\path (257 952)(257 961)
\path (257 952)(257 952)
\path (257 961)(257 961)
\move (257 956) \htext{$\bullet$}
\textref h:C v:C
\path (284 947)(284 961)
\path (284 947)(284 947)
\path (284 961)(284 961)
\move (284 954) \htext{$\bullet$}
\textref h:C v:C
\path (311 941)(311 964)
\path (311 941)(311 941)
\path (311 964)(311 964)
\move (311 952) \htext{$\bullet$}
\textref h:C v:C
\path (338 944)(338 974)
\path (338 944)(338 944)
\path (338 974)(338 974)
\move (338 959) \htext{$\bullet$}
\textref h:C v:C
\path (365 948)(365 974)
\path (365 948)(365 948)
\path (365 974)(365 974)
\move (365 961) \htext{$\bullet$}
\textref h:C v:C
\path (392 919)(392 1002)
\path (392 919)(392 919)
\path (392 1002)(392 1002)
\move (392 960) \htext{$\bullet$}
\textref h:C v:C
\path (419 925)(419 994)
\path (419 925)(419 925)
\path (419 994)(419 994)
\move (419 960) \htext{$\bullet$}
\textref h:C v:C
\path (446 853)(446 1110)
\path (446 853)(446 853)
\path (446 1110)(446 1110)
\move (446 983) \htext{$\bullet$}
\textref h:C v:C
\move (203 834) \htext{$\circ$}
\textref h:C v:C
\move (230 834) \htext{$\circ$}
\textref h:C v:C
\path (257 822)(257 831)
\path (257 822)(257 822)
\path (257 831)(257 831)
\move (257 826) \htext{$\circ$}
\textref h:C v:C
\path (284 827)(284 836)
\path (284 827)(284 827)
\path (284 836)(284 836)
\move (284 831) \htext{$\circ$}
\textref h:C v:C
\path (311 817)(311 833)
\path (311 817)(311 817)
\path (311 833)(311 833)
\move (311 825) \htext{$\circ$}
\textref h:C v:C
\path (338 820)(338 840)
\path (338 820)(338 820)
\path (338 840)(338 840)
\move (338 830) \htext{$\circ$}
\textref h:C v:C
\path (365 824)(365 846)
\path (365 824)(365 824)
\path (365 846)(365 846)
\move (365 835) \htext{$\circ$}
\textref h:C v:C
\path (392 743)(392 836)
\path (392 743)(392 743)
\path (392 836)(392 836)
\move (392 789) \htext{$\circ$}
\textref h:C v:C
\path (419 805)(419 855)
\path (419 805)(419 805)
\path (419 855)(419 855)
\move (419 830) \htext{$\circ$}
\textref h:C v:C
\path (446 710)(446 956)
\path (446 710)(446 710)
\path (446 956)(446 956)
\move (446 830) \htext{$\circ$}
\move(737 625) \textref h:C v:C \htext{}
\move(437 910) \textref h:C v:C \vtext{}
\linewd 3
\path (562 710)(912 710)
\path (562 1110)(912 1110)
\path (562 710)(562 1110)
\path (912 710)(912 1110)
\path (615 710)(615 725)
\path (615 1110)(615 1095)
\path (642 710)(642 725)
\path (642 1110)(642 1095)
\path (669 710)(669 725)
\path (669 1110)(669 1095)
\path (696 710)(696 725)
\path (696 1110)(696 1095)
\path (723 710)(723 725)
\path (723 1110)(723 1095)
\path (750 710)(750 725)
\path (750 1110)(750 1095)
\path (777 710)(777 725)
\path (777 1110)(777 1095)
\path (804 710)(804 725)
\path (804 1110)(804 1095)
\path (831 710)(831 725)
\path (831 1110)(831 1095)
\path (858 710)(858 725)
\path (858 1110)(858 1095)
\textref h:C v:C
\move (615 675) \htext{2}
\move (696 675) \htext{5}
\move (777 675) \htext{8}
\move (858 675) \htext{11}
\path (562 760)(577 760)
\path (912 760)(897 760)
\path (562 810)(577 810)
\path (912 810)(897 810)
\path (562 860)(577 860)
\path (912 860)(897 860)
\path (562 910)(577 910)
\path (912 910)(897 910)
\path (562 960)(577 960)
\path (912 960)(897 960)
\path (562 1010)(577 1010)
\path (912 1010)(897 1010)
\path (562 1060)(577 1060)
\path (912 1060)(897 1060)
\textref h:R v:C
\textref h:C v:C
\move(737 1150) \textref h:C v:C \htext{$R_S$}
\textref h:C v:C
\move (615 760) \htext{$\bullet$}
\textref h:C v:C
\move (642 760) \htext{$\bullet$}
\textref h:C v:C
\move (669 762) \htext{$\bullet$}
\textref h:C v:C
\path (696 754)(696 768)
\path (696 754)(696 754)
\path (696 768)(696 768)
\move (696 761) \htext{$\bullet$}
\textref h:C v:C
\path (723 753)(723 770)
\path (723 753)(723 753)
\path (723 770)(723 770)
\move (723 761) \htext{$\bullet$}
\textref h:C v:C
\path (750 738)(750 768)
\path (750 738)(750 738)
\path (750 768)(750 768)
\move (750 753) \htext{$\bullet$}
\textref h:C v:C
\path (777 750)(777 772)
\path (777 750)(777 750)
\path (777 772)(777 772)
\move (777 761) \htext{$\bullet$}
\textref h:C v:C
\path (804 722)(804 798)
\path (804 722)(804 722)
\path (804 798)(804 798)
\move (804 760) \htext{$\bullet$}
\textref h:C v:C
\path (831 729)(831 782)
\path (831 729)(831 729)
\path (831 782)(831 782)
\move (831 756) \htext{$\bullet$}
\textref h:C v:C
\path (858 710)(858 877)
\path (858 710)(858 710)
\path (858 877)(858 877)
\move (858 748) \htext{$\bullet$}
\textref h:C v:C
\move (615 838) \htext{$\circ$}
\textref h:C v:C
\move (642 840) \htext{$\circ$}
\textref h:C v:C
\move (669 841) \htext{$\circ$}
\textref h:C v:C
\move (696 839) \htext{$\circ$}
\textref h:C v:C
\path (723 832)(723 840)
\path (723 832)(723 832)
\path (723 840)(723 840)
\move (723 836) \htext{$\circ$}
\textref h:C v:C
\path (750 825)(750 841)
\path (750 825)(750 825)
\path (750 841)(750 841)
\move (750 833) \htext{$\circ$}
\textref h:C v:C
\path (777 832)(777 845)
\path (777 832)(777 832)
\path (777 845)(777 845)
\move (777 839) \htext{$\circ$}
\textref h:C v:C
\path (804 832)(804 872)
\path (804 832)(804 832)
\path (804 872)(804 872)
\move (804 852) \htext{$\circ$}
\textref h:C v:C
\path (831 825)(831 853)
\path (831 825)(831 825)
\path (831 853)(831 853)
\move (831 839) \htext{$\circ$}
\textref h:C v:C
\path (858 771)(858 995)
\path (858 771)(858 771)
\path (858 995)(858 995)
\move (858 883) \htext{$\circ$}
\move(1149 625) \textref h:C v:C \htext{}
\move(849 910) \textref h:C v:C \vtext{}
\linewd 3
\path (974 710)(1324 710)
\path (974 1110)(1324 1110)
\path (974 710)(974 1110)
\path (1324 710)(1324 1110)
\path (1027 710)(1027 725)
\path (1027 1110)(1027 1095)
\path (1054 710)(1054 725)
\path (1054 1110)(1054 1095)
\path (1081 710)(1081 725)
\path (1081 1110)(1081 1095)
\path (1108 710)(1108 725)
\path (1108 1110)(1108 1095)
\path (1135 710)(1135 725)
\path (1135 1110)(1135 1095)
\path (1162 710)(1162 725)
\path (1162 1110)(1162 1095)
\path (1189 710)(1189 725)
\path (1189 1110)(1189 1095)
\path (1216 710)(1216 725)
\path (1216 1110)(1216 1095)
\path (1243 710)(1243 725)
\path (1243 1110)(1243 1095)
\path (1270 710)(1270 725)
\path (1270 1110)(1270 1095)
\textref h:C v:C
\move (1027 675) \htext{2}
\move (1108 675) \htext{5}
\move (1189 675) \htext{8}
\move (1270 675) \htext{11}
\path (974 760)(989 760)
\path (1324 760)(1309 760)
\path (974 810)(989 810)
\path (1324 810)(1309 810)
\path (974 860)(989 860)
\path (1324 860)(1309 860)
\path (974 910)(989 910)
\path (1324 910)(1309 910)
\path (974 960)(989 960)
\path (1324 960)(1309 960)
\path (974 1010)(989 1010)
\path (1324 1010)(1309 1010)
\path (974 1060)(989 1060)
\path (1324 1060)(1309 1060)
\textref h:R v:C
\textref h:C v:C
\move(1149 1150) \textref h:C v:C \htext{$R_R$}
\textref h:C v:C
\move (1027 962) \htext{$\bullet$}
\textref h:C v:C
\move (1054 962) \htext{$\bullet$}
\textref h:C v:C
\path (1081 955)(1081 968)
\path (1081 955)(1081 955)
\path (1081 968)(1081 968)
\move (1081 961) \htext{$\bullet$}
\textref h:C v:C
\path (1108 956)(1108 969)
\path (1108 956)(1108 956)
\path (1108 969)(1108 969)
\move (1108 962) \htext{$\bullet$}
\textref h:C v:C
\path (1135 950)(1135 975)
\path (1135 950)(1135 950)
\path (1135 975)(1135 975)
\move (1135 962) \htext{$\bullet$}
\textref h:C v:C
\path (1162 943)(1162 974)
\path (1162 943)(1162 943)
\path (1162 974)(1162 974)
\move (1162 958) \htext{$\bullet$}
\textref h:C v:C
\path (1189 939)(1189 987)
\path (1189 939)(1189 939)
\path (1189 987)(1189 987)
\move (1189 963) \htext{$\bullet$}
\textref h:C v:C
\path (1216 972)(1216 1090)
\path (1216 972)(1216 972)
\path (1216 1090)(1216 1090)
\move (1216 1031) \htext{$\bullet$}
\textref h:C v:C
\path (1243 937)(1243 995)
\path (1243 937)(1243 937)
\path (1243 995)(1243 995)
\move (1243 966) \htext{$\bullet$}
\textref h:C v:C
\path (1270 819)(1270 1110)
\path (1270 819)(1270 819)
\path (1270 1110)(1270 1110)
\move (1270 977) \htext{$\bullet$}
\textref h:C v:C
\move (1027 831) \htext{$\circ$}
\textref h:C v:C
\move (1054 831) \htext{$\circ$}
\textref h:C v:C
\path (1081 822)(1081 833)
\path (1081 822)(1081 822)
\path (1081 833)(1081 833)
\move (1081 828) \htext{$\circ$}
\textref h:C v:C
\path (1108 815)(1108 833)
\path (1108 815)(1108 815)
\path (1108 833)(1108 833)
\move (1108 824) \htext{$\circ$}
\textref h:C v:C
\path (1135 820)(1135 837)
\path (1135 820)(1135 820)
\path (1135 837)(1135 837)
\move (1135 829) \htext{$\circ$}
\textref h:C v:C
\path (1162 785)(1162 829)
\path (1162 785)(1162 785)
\path (1162 829)(1162 829)
\move (1162 807) \htext{$\circ$}
\textref h:C v:C
\path (1189 735)(1189 842)
\path (1189 735)(1189 735)
\path (1189 842)(1189 842)
\move (1189 789) \htext{$\circ$}
\textref h:C v:C
\path (1216 767)(1216 851)
\path (1216 767)(1216 767)
\path (1216 851)(1216 851)
\move (1216 809) \htext{$\circ$}
\textref h:C v:C
\path (1243 771)(1243 832)
\path (1243 771)(1243 771)
\path (1243 832)(1243 832)
\move (1243 801) \htext{$\circ$}
\textref h:C v:C
\path (1270 773)(1270 955)
\path (1270 773)(1270 773)
\path (1270 955)(1270 955)
\move (1270 864) \htext{$\circ$}
\move(1561 625) \textref h:C v:C \htext{}
\move(1261 910) \textref h:C v:C \vtext{}
\linewd 3
\path (1386 710)(1736 710)
\path (1386 1110)(1736 1110)
\path (1386 710)(1386 1110)
\path (1736 710)(1736 1110)
\path (1439 710)(1439 725)
\path (1439 1110)(1439 1095)
\path (1466 710)(1466 725)
\path (1466 1110)(1466 1095)
\path (1493 710)(1493 725)
\path (1493 1110)(1493 1095)
\path (1520 710)(1520 725)
\path (1520 1110)(1520 1095)
\path (1547 710)(1547 725)
\path (1547 1110)(1547 1095)
\path (1574 710)(1574 725)
\path (1574 1110)(1574 1095)
\path (1601 710)(1601 725)
\path (1601 1110)(1601 1095)
\path (1628 710)(1628 725)
\path (1628 1110)(1628 1095)
\path (1655 710)(1655 725)
\path (1655 1110)(1655 1095)
\path (1682 710)(1682 725)
\path (1682 1110)(1682 1095)
\textref h:C v:C
\move (1439 675) \htext{2}
\move (1520 675) \htext{5}
\move (1601 675) \htext{8}
\move (1682 675) \htext{11}
\path (1386 760)(1401 760)
\path (1736 760)(1721 760)
\path (1386 810)(1401 810)
\path (1736 810)(1721 810)
\path (1386 860)(1401 860)
\path (1736 860)(1721 860)
\path (1386 910)(1401 910)
\path (1736 910)(1721 910)
\path (1386 960)(1401 960)
\path (1736 960)(1721 960)
\path (1386 1010)(1401 1010)
\path (1736 1010)(1721 1010)
\path (1386 1060)(1401 1060)
\path (1736 1060)(1721 1060)
\textref h:R v:C
\textref h:C v:C
\move(1561 1150) \textref h:C v:C \htext{$R_N$}
\textref h:C v:C
\path (1439 1060)(1439 1070)
\path (1439 1060)(1439 1060)
\path (1439 1070)(1439 1070)
\move (1439 1065) \htext{$\bullet$}
\textref h:C v:C
\path (1466 1048)(1466 1063)
\path (1466 1048)(1466 1048)
\path (1466 1063)(1466 1063)
\move (1466 1056) \htext{$\bullet$}
\textref h:C v:C
\path (1493 1050)(1493 1074)
\path (1493 1050)(1493 1050)
\path (1493 1074)(1493 1074)
\move (1493 1062) \htext{$\bullet$}
\textref h:C v:C
\path (1520 1047)(1520 1084)
\path (1520 1047)(1520 1047)
\path (1520 1084)(1520 1084)
\move (1520 1065) \htext{$\bullet$}
\textref h:C v:C
\path (1547 1052)(1547 1081)
\path (1547 1052)(1547 1052)
\path (1547 1081)(1547 1081)
\move (1547 1067) \htext{$\bullet$}
\textref h:C v:C
\path (1574 1073)(1574 1110)
\path (1574 1073)(1574 1073)
\path (1574 1110)(1574 1110)
\move (1574 1096) \htext{$\bullet$}
\textref h:C v:C
\path (1601 1052)(1601 1105)
\path (1601 1052)(1601 1052)
\path (1601 1105)(1601 1105)
\move (1601 1078) \htext{$\bullet$}
\textref h:C v:C
\path (1628 957)(1628 1110)
\path (1628 957)(1628 957)
\path (1628 1110)(1628 1110)
\move (1628 1056) \htext{$\bullet$}
\textref h:C v:C
\path (1655 1026)(1655 1110)
\path (1655 1026)(1655 1026)
\path (1655 1110)(1655 1110)
\move (1655 1083) \htext{$\bullet$}
\textref h:C v:C
\path (1682 794)(1682 1110)
\path (1682 794)(1682 794)
\path (1682 1110)(1682 1110)
\textref h:C v:C
\path (1439 929)(1439 937)
\path (1439 929)(1439 929)
\path (1439 937)(1439 937)
\move (1439 933) \htext{$\circ$}
\textref h:C v:C
\path (1466 919)(1466 934)
\path (1466 919)(1466 919)
\path (1466 934)(1466 934)
\move (1466 926) \htext{$\circ$}
\textref h:C v:C
\path (1493 921)(1493 940)
\path (1493 921)(1493 921)
\path (1493 940)(1493 940)
\move (1493 930) \htext{$\circ$}
\textref h:C v:C
\path (1520 913)(1520 941)
\path (1520 913)(1520 913)
\path (1520 941)(1520 941)
\move (1520 927) \htext{$\circ$}
\textref h:C v:C
\path (1547 912)(1547 940)
\path (1547 912)(1547 912)
\path (1547 940)(1547 940)
\move (1547 926) \htext{$\circ$}
\textref h:C v:C
\path (1574 904)(1574 975)
\path (1574 904)(1574 904)
\path (1574 975)(1574 975)
\move (1574 939) \htext{$\circ$}
\textref h:C v:C
\path (1601 871)(1601 956)
\path (1601 871)(1601 871)
\path (1601 956)(1601 956)
\move (1601 914) \htext{$\circ$}
\textref h:C v:C
\path (1628 875)(1628 1039)
\path (1628 875)(1628 875)
\path (1628 1039)(1628 1039)
\move (1628 957) \htext{$\circ$}
\textref h:C v:C
\path (1655 928)(1655 1007)
\path (1655 928)(1655 928)
\path (1655 1007)(1655 1007)
\move (1655 968) \htext{$\circ$}
\textref h:C v:C
\path (1682 755)(1682 1110)
\path (1682 755)(1682 755)
\path (1682 1110)(1682 1110)
\move (1682 979) \htext{$\circ$}
\move(950 25) \textref h:C v:C \htext{distance $t_2$ $(t_1=3)$}
\move(25 310) \textref h:C v:C \vtext{$R_i=\frac{3}{8}(2 R_i^a-2 R_i^b)$}
\linewd 3
\path (150 110)(150 510)
\textref h:C v:C
\path (150 176)(165 176)
\path (150 243)(165 243)
\path (150 310)(165 310)
\path (150 376)(165 376)
\path (150 443)(165 443)
\textref h:R v:C
\move (135 176) \htext{-1.0}
\move (135 310) \htext{0.0}
\move (135 443) \htext{1.0}
\textref h:C v:C
\move(325 25) \textref h:C v:C \htext{}
\move(25 310) \textref h:C v:C \vtext{}
\linewd 3
\path (150 110)(500 110)
\path (150 510)(500 510)
\path (150 110)(150 510)
\path (500 110)(500 510)
\path (203 110)(203 125)
\path (203 510)(203 495)
\path (230 110)(230 125)
\path (230 510)(230 495)
\path (257 110)(257 125)
\path (257 510)(257 495)
\path (284 110)(284 125)
\path (284 510)(284 495)
\path (311 110)(311 125)
\path (311 510)(311 495)
\path (338 110)(338 125)
\path (338 510)(338 495)
\path (365 110)(365 125)
\path (365 510)(365 495)
\path (392 110)(392 125)
\path (392 510)(392 495)
\path (419 110)(419 125)
\path (419 510)(419 495)
\path (446 110)(446 125)
\path (446 510)(446 495)
\textref h:C v:C
\move (203 75) \htext{2}
\move (284 75) \htext{5}
\move (365 75) \htext{8}
\move (446 75) \htext{11}
\path (150 176)(165 176)
\path (500 176)(485 176)
\path (150 243)(165 243)
\path (500 243)(485 243)
\path (150 310)(165 310)
\path (500 310)(485 310)
\path (150 376)(165 376)
\path (500 376)(485 376)
\path (150 443)(165 443)
\path (500 443)(485 443)
\textref h:R v:C
\textref h:C v:C
\textref h:C v:C
\move (203 434) \htext{$\bullet$}
\textref h:C v:C
\path (230 428)(230 439)
\path (230 428)(230 428)
\path (230 439)(230 439)
\move (230 434) \htext{$\bullet$}
\textref h:C v:C
\path (257 433)(257 446)
\path (257 433)(257 433)
\path (257 446)(257 446)
\move (257 440) \htext{$\bullet$}
\textref h:C v:C
\path (284 424)(284 441)
\path (284 424)(284 424)
\path (284 441)(284 441)
\move (284 432) \htext{$\bullet$}
\textref h:C v:C
\path (311 422)(311 450)
\path (311 422)(311 422)
\path (311 450)(311 450)
\move (311 436) \htext{$\bullet$}
\textref h:C v:C
\path (338 420)(338 457)
\path (338 420)(338 420)
\path (338 457)(338 457)
\move (338 439) \htext{$\bullet$}
\textref h:C v:C
\path (365 419)(365 453)
\path (365 419)(365 419)
\path (365 453)(365 453)
\move (365 436) \htext{$\bullet$}
\textref h:C v:C
\path (392 418)(392 510)
\path (392 418)(392 418)
\path (392 510)(392 510)
\move (392 480) \htext{$\bullet$}
\textref h:C v:C
\path (419 397)(419 482)
\path (419 397)(419 397)
\path (419 482)(419 482)
\move (419 439) \htext{$\bullet$}
\textref h:C v:C
\path (446 281)(446 510)
\path (446 281)(446 281)
\path (446 510)(446 510)
\move (446 462) \htext{$\bullet$}
\move(737 25) \textref h:C v:C \htext{}
\move(437 310) \textref h:C v:C \vtext{}
\linewd 3
\path (562 110)(912 110)
\path (562 510)(912 510)
\path (562 110)(562 510)
\path (912 110)(912 510)
\path (615 110)(615 125)
\path (615 510)(615 495)
\path (642 110)(642 125)
\path (642 510)(642 495)
\path (669 110)(669 125)
\path (669 510)(669 495)
\path (696 110)(696 125)
\path (696 510)(696 495)
\path (723 110)(723 125)
\path (723 510)(723 495)
\path (750 110)(750 125)
\path (750 510)(750 495)
\path (777 110)(777 125)
\path (777 510)(777 495)
\path (804 110)(804 125)
\path (804 510)(804 495)
\path (831 110)(831 125)
\path (831 510)(831 495)
\path (858 110)(858 125)
\path (858 510)(858 495)
\textref h:C v:C
\move (615 75) \htext{2}
\move (696 75) \htext{5}
\move (777 75) \htext{8}
\move (858 75) \htext{11}
\path (562 176)(577 176)
\path (912 176)(897 176)
\path (562 243)(577 243)
\path (912 243)(897 243)
\path (562 310)(577 310)
\path (912 310)(897 310)
\path (562 376)(577 376)
\path (912 376)(897 376)
\path (562 443)(577 443)
\path (912 443)(897 443)
\textref h:R v:C
\textref h:C v:C
\textref h:C v:C
\move (615 232) \htext{$\bullet$}
\textref h:C v:C
\move (642 229) \htext{$\bullet$}
\textref h:C v:C
\move (669 231) \htext{$\bullet$}
\textref h:C v:C
\path (696 224)(696 239)
\path (696 224)(696 224)
\path (696 239)(696 239)
\move (696 231) \htext{$\bullet$}
\textref h:C v:C
\path (723 225)(723 244)
\path (723 225)(723 225)
\path (723 244)(723 244)
\move (723 235) \htext{$\bullet$}
\textref h:C v:C
\path (750 213)(750 246)
\path (750 213)(750 213)
\path (750 246)(750 246)
\move (750 230) \htext{$\bullet$}
\textref h:C v:C
\path (777 219)(777 245)
\path (777 219)(777 219)
\path (777 245)(777 245)
\move (777 232) \htext{$\bullet$}
\textref h:C v:C
\path (804 175)(804 260)
\path (804 175)(804 175)
\path (804 260)(804 260)
\move (804 217) \htext{$\bullet$}
\textref h:C v:C
\path (831 196)(831 256)
\path (831 196)(831 196)
\path (831 256)(831 256)
\move (831 226) \htext{$\bullet$}
\textref h:C v:C
\path (858 110)(858 345)
\path (858 110)(858 110)
\path (858 345)(858 345)
\move (858 175) \htext{$\bullet$}
\move(1149 25) \textref h:C v:C \htext{}
\move(849 310) \textref h:C v:C \vtext{}
\linewd 3
\path (974 110)(1324 110)
\path (974 510)(1324 510)
\path (974 110)(974 510)
\path (1324 110)(1324 510)
\path (1027 110)(1027 125)
\path (1027 510)(1027 495)
\path (1054 110)(1054 125)
\path (1054 510)(1054 495)
\path (1081 110)(1081 125)
\path (1081 510)(1081 495)
\path (1108 110)(1108 125)
\path (1108 510)(1108 495)
\path (1135 110)(1135 125)
\path (1135 510)(1135 495)
\path (1162 110)(1162 125)
\path (1162 510)(1162 495)
\path (1189 110)(1189 125)
\path (1189 510)(1189 495)
\path (1216 110)(1216 125)
\path (1216 510)(1216 495)
\path (1243 110)(1243 125)
\path (1243 510)(1243 495)
\path (1270 110)(1270 125)
\path (1270 510)(1270 495)
\textref h:C v:C
\move (1027 75) \htext{2}
\move (1108 75) \htext{5}
\move (1189 75) \htext{8}
\move (1270 75) \htext{11}
\path (974 176)(989 176)
\path (1324 176)(1309 176)
\path (974 243)(989 243)
\path (1324 243)(1309 243)
\path (974 310)(989 310)
\path (1324 310)(1309 310)
\path (974 376)(989 376)
\path (1324 376)(1309 376)
\path (974 443)(989 443)
\path (1324 443)(1309 443)
\textref h:R v:C
\textref h:C v:C
\textref h:C v:C
\move (1027 440) \htext{$\bullet$}
\textref h:C v:C
\move (1054 440) \htext{$\bullet$}
\textref h:C v:C
\path (1081 435)(1081 452)
\path (1081 435)(1081 435)
\path (1081 452)(1081 452)
\move (1081 443) \htext{$\bullet$}
\textref h:C v:C
\path (1108 437)(1108 459)
\path (1108 437)(1108 437)
\path (1108 459)(1108 459)
\move (1108 448) \htext{$\bullet$}
\textref h:C v:C
\path (1135 428)(1135 458)
\path (1135 428)(1135 428)
\path (1135 458)(1135 458)
\move (1135 443) \htext{$\bullet$}
\textref h:C v:C
\path (1162 434)(1162 488)
\path (1162 434)(1162 434)
\path (1162 488)(1162 488)
\move (1162 461) \htext{$\bullet$}
\textref h:C v:C
\path (1189 425)(1189 510)
\path (1189 425)(1189 425)
\path (1189 510)(1189 510)
\move (1189 484) \htext{$\bullet$}
\textref h:C v:C
\path (1216 459)(1216 510)
\path (1216 459)(1216 459)
\path (1216 510)(1216 510)
\textref h:C v:C
\path (1243 432)(1243 510)
\path (1243 432)(1243 432)
\path (1243 510)(1243 510)
\move (1243 474) \htext{$\bullet$}
\textref h:C v:C
\path (1270 239)(1270 510)
\path (1270 239)(1270 239)
\path (1270 510)(1270 510)
\move (1270 422) \htext{$\bullet$}
\move(1561 25) \textref h:C v:C \htext{}
\move(1261 310) \textref h:C v:C \vtext{}
\linewd 3
\path (1386 110)(1736 110)
\path (1386 510)(1736 510)
\path (1386 110)(1386 510)
\path (1736 110)(1736 510)
\path (1439 110)(1439 125)
\path (1439 510)(1439 495)
\path (1466 110)(1466 125)
\path (1466 510)(1466 495)
\path (1493 110)(1493 125)
\path (1493 510)(1493 495)
\path (1520 110)(1520 125)
\path (1520 510)(1520 495)
\path (1547 110)(1547 125)
\path (1547 510)(1547 495)
\path (1574 110)(1574 125)
\path (1574 510)(1574 495)
\path (1601 110)(1601 125)
\path (1601 510)(1601 495)
\path (1628 110)(1628 125)
\path (1628 510)(1628 495)
\path (1655 110)(1655 125)
\path (1655 510)(1655 495)
\path (1682 110)(1682 125)
\path (1682 510)(1682 495)
\textref h:C v:C
\move (1439 75) \htext{2}
\move (1520 75) \htext{5}
\move (1601 75) \htext{8}
\move (1682 75) \htext{11}
\path (1386 176)(1401 176)
\path (1736 176)(1721 176)
\path (1386 243)(1401 243)
\path (1736 243)(1721 243)
\path (1386 310)(1401 310)
\path (1736 310)(1721 310)
\path (1386 376)(1401 376)
\path (1736 376)(1721 376)
\path (1386 443)(1401 443)
\path (1736 443)(1721 443)
\textref h:R v:C
\textref h:C v:C
\textref h:C v:C
\path (1439 434)(1439 447)
\path (1439 434)(1439 434)
\path (1439 447)(1439 447)
\move (1439 441) \htext{$\bullet$}
\textref h:C v:C
\path (1466 428)(1466 449)
\path (1466 428)(1466 428)
\path (1466 449)(1466 449)
\move (1466 439) \htext{$\bullet$}
\textref h:C v:C
\path (1493 426)(1493 456)
\path (1493 426)(1493 426)
\path (1493 456)(1493 456)
\move (1493 441) \htext{$\bullet$}
\textref h:C v:C
\path (1520 424)(1520 471)
\path (1520 424)(1520 424)
\path (1520 471)(1520 471)
\move (1520 448) \htext{$\bullet$}
\textref h:C v:C
\path (1547 430)(1547 470)
\path (1547 430)(1547 430)
\path (1547 470)(1547 470)
\move (1547 450) \htext{$\bullet$}
\textref h:C v:C
\path (1574 424)(1574 508)
\path (1574 424)(1574 424)
\path (1574 508)(1574 508)
\move (1574 466) \htext{$\bullet$}
\textref h:C v:C
\path (1601 424)(1601 510)
\path (1601 424)(1601 424)
\path (1601 510)(1601 510)
\move (1601 474) \htext{$\bullet$}
\textref h:C v:C
\path (1628 280)(1628 510)
\path (1628 280)(1628 280)
\path (1628 510)(1628 510)
\move (1628 408) \htext{$\bullet$}
\textref h:C v:C
\path (1655 355)(1655 495)
\path (1655 355)(1655 355)
\path (1655 495)(1655 495)
\move (1655 425) \htext{$\bullet$}
\textref h:C v:C
\path (1682 110)(1682 510)
\path (1682 110)(1682 110)
\path (1682 510)(1682 510)
\end{texdraw}}
\end{center}